\documentclass[12pt,a4paper]{article}

\usepackage{amsmath, amssymb}

\usepackage{graphicx}

\usepackage[english]{babel}

\usepackage{a4wide}

\usepackage{xcolor}

\usepackage[hidelinks]{hyperref}
\usepackage{url}

\usepackage{enumitem}

\usepackage{subcaption}

\usepackage{float}

\usepackage{siunitx}
\usepackage{epstopdf}

\usepackage{authblk}

\allowdisplaybreaks

\newcommand{\bs}{\boldsymbol}

\newcommand{\bTheta}{\ensuremath{\bs{\Theta}}}

\newcommand{\ionenotrunc}{\ensuremath{t^l \leq l_i=u_i\leq t}}
\newcommand{\itwonotrunc}{\ensuremath{t^l<t< l_i=u_i}}
\newcommand{\ithreenotrunc}{\ensuremath{t^l\leq l_i<u_i\leq t}}
\newcommand{\ifournotrunc}{\ensuremath{t^l< t \leq l_i<u_i}}
\newcommand{\ifivenotrunc}{\ensuremath{t^l\leq l_i < t<u_i}}
\newcommand{\ifivelowernotrunc}{\ensuremath{t^l \leq l_i<X_i \leq t<u_i}}
\newcommand{\ifiveuppernotrunc}{\ensuremath{t^l \leq l_i\leq t<X_i<u_i}}

\newcommand{\ione}{\ensuremath{\ionenotrunc < T}}
\newcommand{\itwo}{\ensuremath{\itwonotrunc \leq T}}
\newcommand{\ithree}{\ensuremath{\ithreenotrunc<T}}
\newcommand{\ifour}{\ensuremath{\ifournotrunc \leq T}}
\newcommand{\ifive}{\ensuremath{\ifivenotrunc \leq T}}
\newcommand{\ifivelower}{\ensuremath{\ifivelowernotrunc \leq T}}
\newcommand{\ifiveupper}{\ensuremath{\ifiveuppernotrunc \leq T}}

\DeclareMathOperator{\VaR}{VaR}
\DeclareMathOperator{\TVaR}{TVaR}

\newcommand*{\cond}{\,\ifnum\currentgrouptype=16 \middle\fi|\,}

\providecommand{\keywords}[1]{\textbf{Keywords:} #1}

\usepackage{natbib}
\title{Modelling censored losses using splicing:\\ a global fit strategy with mixed Erlang and extreme value distributions}

\author[a]{Tom Reynkens\thanks{Corresponding author. \\ \textit{Email addresses:} \href{mailto:tom.reynkens@kuleuven.be}{tom.reynkens@kuleuven.be} (T.~Reynkens), \href{mailto:roel.verbelen@kuleuven.be}{roel.verbelen@kuleuven.be} (R.~Verbelen), \href{mailto:jan.beirlant@kuleuven.be}{jan.beirlant@kuleuven.be}  (J.~Beirlant) and \href{mailto:katrien.antonio@kuleuven.be}{katrien.antonio@kuleuven.be} (K.~Antonio).}}
\author[b]{Roel Verbelen}
\author[a,c]{Jan Beirlant}
\author[b,d]{Katrien Antonio}
\affil[a]{LStat and LRisk, Department of Mathematics, KU Leuven. \textit{Celestijnenlaan 200B, 3001 Leuven, Belgium.}}
\affil[b]{LStat and LRisk, Faculty of Economics and Business, KU Leuven. \textit{Naamsestraat 69, 3000 Leuven, Belgium.}}
\affil[c]{Department of Mathematical Statistics and Actuarial Science, University of the Free State. \textit{P.O. Box 339, Bloemfontein 9300, South Africa.}}
\affil[d]{Faculty of Economics and Business, University of Amsterdam. \textit{Roetersstraat 11, 1018 WB Amsterdam, The Netherlands.}}

\begin{document}
\maketitle

\begin{abstract}
\noindent In risk analysis, a global fit that appropriately captures the body and the tail of the distribution of losses is essential. Modelling the whole range of the losses using a standard distribution is usually very hard and often impossible due to the specific characteristics of the body and the tail of the loss distribution. A possible solution is to combine two distributions in a splicing model: a light-tailed distribution for the body which covers light and moderate losses, and a heavy-tailed distribution for the tail to capture large losses. We propose a splicing model with a mixed Erlang (ME) distribution for the body and a Pareto distribution for the tail. This combines the flexibility of the ME distribution with the ability of the Pareto distribution to model extreme values. We extend our splicing approach for censored and/or truncated data. Relevant examples of such data can be found in financial risk analysis. We illustrate the flexibility of this splicing model using practical examples from risk measurement.
\end{abstract}

\section{Introduction}\label{sec:intro}

In several domains such as insurance, finance and operational risk, modelling financial losses is essential. For example, actuaries use models for claim sizes to set premiums, calculate risk measures and determine capital requirements for solvency regulations. This type of data is typically heavy-tailed and high losses can occur. A standard parametric distribution for the tail is a Pareto-type distribution, which is of key importance in extreme value theory \citep[see e.g.][]{McNeil_EVT}. The Pareto distribution or the generalised Pareto distribution (GPD) are used to model exceedances over intermediate thresholds. However, they are not able to capture the characteristics over the whole range of the loss distribution which makes them not suitable as a global fit distribution (see e.g.~\citet{McNeil_EVT}, Section~6.4 in \citet{Embrechts} and Section~6.2 in \citet{SoE}). It is often imperative to obtain a global fit for the distribution of losses, for example in a risk analysis where focus is not only on extreme events, or when setting up a reinsurance program.
Instead of trying many different standard distributions, splicing two distributions \citep{LossModels} is more suitable to model the complete loss distribution. In literature, a splicing model is also called a composite model. We hereby combine a light-tailed distribution for the body which covers light and moderate losses (the so-called attritional losses), and a heavy-tailed distribution for the tail to capture large losses. In the actuarial literature simple splicing models have been proposed. \citet{SoE, LossModels} consider the splicing of the exponential distribution with the Pareto distribution. Other distributions for the body such as the Weibull distribution \citep{Weibull_Pa, Weibull_Pa2} or the log-normal distribution \citep{LN_Pa, LN_Pa2, LN_Pa3} have also been used. \citet{comp_LN, comp_Weibull, Stoppa} investigate the splicing of the log-normal or Weibull distribution with various tail distributions. \citet{EM_exp_GPD} consider the splicing of a mixture of two exponentials and the GPD. The use of a mixture model in the first splicing component gives more flexibility in modelling the light and moderate losses. \citet{Fackler} provides an overview of spliced distributions for loss modelling, and \citet{BrazKlee} illustrate the modelling performance of several spliced distributions on a real data example. Note that splicing has not only been considered in an actuarial context. \citet{Panjer,Deutsche, OR_tails} use this technique to model operational risk data.

\vspace*{0.25\baselineskip}
 \par The mixed Erlang (ME) distribution became popular in loss modelling because of several reasons \citep[see e.g.][]{WillmotWoo,LeeLin,WillmotLin,LossModels_FT}. The class of ME distributions with common scale parameter is dense in the space of positive continuous distributions \citep{Tijms}. Any positive continuous distribution can thus be approximated up to any given accuracy by a ME distribution. This class is also closed under mixture, convolution and compounding. Therefore, we can readily obtain aggregate loss distributions removing the need for simulations. Moreover, we can easily compute risk measures such as the Value-at-Risk (VaR), the Tail VaR (TVaR) and premiums of excess-loss insurances.

	\par Fitting the ME distribution using direct likelihood maximisation is difficult.  The ME parameters can also be estimated based on the denseness proof of \citet{Tijms} but this method converges slowly and leads to overfitting \citep{LeeLin}. The preferred strategy is to use the expectation-maximisation (EM) algorithm \citep{EM} to fit the ME distribution as proposed by \citet{LeeLin}. An advantage is that the E- and M-steps can be solved analytically.
\citet{LeeLin} use information criteria (IC) like the Akaike information criterion (AIC, \cite{AIC}) or the Bayesian information criterion (BIC, \cite{BIC}) to select the number of components in the mixture and as such avoid overfitting. 

\vspace*{0.25\baselineskip}
\par Our work is further inspired by the omnipresence of censoring and truncation in risk analysis and risk modelling, see e.g.~\citet{creditrisk, LossModels, AntonioPlat, ME}. 

\par \textit{Lower truncation} occurs when payments that are below certain thresholds are not observed. In insurance, lower truncation occurs, for example, due to the presence of a deductible in the insurance contract. In some practical applications, there might be a natural bound that upper truncates the tail distribution. For example, earthquake magnitudes, forest fire areas and daily price changes of stocks have distributions that are naturally \textit{upper truncated} \citep{trHill}. In an insurance context, where premiums have to be set using the fitted model, introducing an upper truncation point can prevent probability mass being assigned to unreasonably large claim amounts.

\par \textit{Right censoring} is highly relevant in the context of loss models and risk measurement for unsettled claims in non-life insurance and reinsurance. The (re)insurer only knows the true cost of a policy when all claims on this policy are settled or closed. However, in the development or lifetime of a non-life insurance claim, a significant time may elapse between the claim occurrence and its final settlement or closure. For such unsettled claims only the payment to date is known and the quantity of interest, i.e.~the final cumulative payment on a claim, is right censored. This complicates the calculation of reinsurance premiums for large claims and forces the insurer to predict, with maximum accuracy, the capital buffer that is required to indemnify the insured in the future regarding claims that happened in the past. To support this complex task, actuaries will use additional, expert information called \textit{incurred data}. This is the sum of the actual payment (so far) on a claim and its case estimate. These case estimates are set by an experienced case handler and express the expert's estimate of the outstanding loss on a claim. For large claims, facing very long settlement (e.g.~due to legal procedures or severe bodily injury), actuaries consider incurred data as a highly important source of information. We propose to construct upper bounds for the final cumulative payment on a claim using incurreds. When the true final claim amount lies between the cumulative payment up to date and the incurred value, \textit{interval censoring} techniques can be applied as we will demonstrate in this paper.

\par In the splicing context, some work has already been done for censored and/or truncated data. \citet{TruncWeibull_Pa} take lower truncation into account for Weibull-Pareto splicing.
\citet{ME} extend the mixed Erlang approach of \citet{LeeLin} to censored and/or truncated data. \citet{cEVT1, cEVT2} discuss extensions of classical extreme value estimators to right censored data. Extensions to upper truncated data have been investigated by \citet{trHill,Truncation}.

\vspace*{0.25\baselineskip}
\par Although the ME distribution has several advantages, as discussed above, one major disadvantage is that it has an asymptotically exponential, and hence:~light, tail \citep{Neuts}.
Therefore, overfitting can still occur on heavy-tailed data as one needs many components to model the heavy-tailedness appropriately. The simulated sample of the GPD in \citet{ME} illustrates this behaviour. As a first contribution, we overcome this drawback by proposing a splicing model with the ME distribution for the body and the Pareto distribution for the tail (Section~\ref{sec:splicing}). A global fit for financial loss data then results, which combines the flexibility of the ME distribution to model light and moderate losses with the ability of the Pareto distribution to model heavy-tailed data. Fire and motor third party liability (MTPL) insurance losses, and financial returns are examples of heavy-tailed data which are of Pareto type. This strategy avoids ad hoc combinations of a standard light-tailed distribution, such as the log-normal or the Weibull distribution, for the body with a heavy-tailed distribution for the tail, as explored in many papers on loss modelling. Moreover, a mixture of Erlangs yields more flexibility than a mixture of two exponentials as in \citet{EM_exp_GPD} while keeping analytic tractability.

\par As a second contribution, we extend the global fit strategy based on splicing to take both (random) censoring and truncation into account. Up to our knowledge, this full framework has not yet been considered in the literature. We provide a general fitting procedure for the model using the EM algorithm where the incompleteness is caused by censoring, see Section~\ref{sec:EMcens}.
Instead of using a splicing model, a common technique in extreme value analysis is to combine a non-parametric fit for the body and a parametric model (e.g.~Pareto distribution) for the tail. However, when censoring is present, this approach can no longer be applied as we might have interval censored data points where the lower bound of the interval is in the body of the distribution, whereas the upper bound is in the tail. Our general splicing framework can handle observations of this type and can hence be used to provide a global fit. In Section~\ref{sec:ME_Par}, we apply the general fitting procedure for censored and/or truncated data to the specific case of our ME-Pareto splicing model. As we provide a general procedure to fit a splicing model to censored and/or truncated data, we could possibly use another extreme value distribution, such as the GPD, 
 instead of the Pareto distribution. For the GPD, however, in case there is censoring, the expectations in the E-step can no longer be computed analytically, in contrast to the Pareto distribution.

\par Finally, we discuss the computation of risk measures using our splicing model in Section~\ref{sec:riskmeasures} and we apply the method to two real life data examples in Section~\ref{sec:examples}.

\section{Splicing of ME and Pareto distributions}\label{sec:splicing}
\subsection{General splicing model}\label{sec:gensplicing}

Consider two densities $f_1^*$ and $f_2^*$, and denote the corresponding cumulative distribution functions (CDFs) by $F_1^*$ and $F_2^*$. Their parameters are
contained in the vectors $\bTheta_1$ and $\bTheta_2$, respectively. We assume that there are no shared parameters in $\bTheta_1$ and $\bTheta_2$. Define now 
\[f_1(x; t^l,t,\bTheta_1) = \begin{cases}
\frac{f_1^*(x; \, \bTheta_1)}{F_1^*(t; \,\bTheta_1)-F_1^*(t^l; \,\bTheta_1)} & \text{if } t^l\leq x \leq t\\
0 & \text{otherwise},
\end{cases}
\]
\[ f_2(x; t,T,\bTheta_2) = \begin{cases}
\frac{f_2^*(x; \,\bTheta_2)}{F_2^*(T; \,\bTheta_2)-F_2^*(t; \,\bTheta_2)} & \text{if } t\leq x\leq T \\
0 & \text{otherwise},
\end{cases}
\]
\newline where $0\leq t^l<t < T$ are fixed points.
The first density is lower truncated at $t^l$ and upper truncated at $t$, and the second density is lower truncated at $t$ and upper truncated at $T$.
The density for the body, $f_1$, and density for the tail, $f_2$, are then valid densities on the intervals $[t^l,t]$ and $[t,T]$, respectively.
In case of no upper truncation for the tail distribution, we set $T = + \infty$.
The corresponding CDFs are 
\[F_1(x; t^l,t,\bTheta_1) = \begin{cases}
0 & \text{if } x\leq t^l \\
\frac{F_1^*(x; \,\bTheta_1)-F_1^*(t^l; \,\bTheta_1)}{F_1^*(t; \,\bTheta_1)-F_1^*(t^l; \,\bTheta_1)} & \text{if } t^l< x < t\\
1 & \text{if } x\geq t,
\end{cases}
\]
\[ F_2(x;t,T,\bTheta_2) = \begin{cases}
0 & \text{if } x\leq t\\
\frac{F_2^*(x; \,\bTheta_2)-F_2^*(t; \,\bTheta_2)}{F_2^*(T; \,\bTheta_2)-F_2^*(t; \,\bTheta_2)} & \text{if } t<x<T \\
1 & \text{if } x\geq T.
\end{cases}
\]
\par Consider the splicing weight $\pi \in (0,1)$. The spliced density is then defined as
\[f(x;t^l,t,T,\bTheta) = \begin{cases}
0 & \quad \text{if } x\leq t^l\\
\pi f_1(x; t^l,t,\bTheta_1) & \quad \text{if }t^l<x\leq t \\
(1-\pi) f_2(x;t,T,\bTheta_2) &\quad \text{if } t<x < T\\
0 & \quad \text{if } x\geq T,
\end{cases}
\]
where $\bTheta=(\pi,\bTheta_1,\bTheta_2)$ is the parameter vector. We call the point $t$ the splicing point, and the points $t^l$ and $T$ the lower, respectively, upper truncation points. The corresponding, continuous, CDF is given by
\begin{equation}
F(x;t^l,t,T,\bTheta) = \begin{cases}
0 & \quad \text{if }  x\leq t^l\\
\pi F_1(x; t^l,t,\bTheta_1) & \quad \text{if } t^l<x\leq t \\
\pi+(1-\pi) F_2(x;t,T,\bTheta_2) &\quad \text{if } t<x< T \\
1 &\quad \text{if } x\geq T.
\end{cases}
\label{eq:spliceCDF}
\end{equation}
Most authors impose differentiability of the probability density function (PDF) at the splicing point to get a smooth density function and to reduce the number of parameters. The splicing point is then estimated together with the other model parameters using maximum likelihood estimation (MLE). This restriction results in less flexibility. Therefore, we choose to not follow this approach, but determine the splicing point directly using an extreme value analysis, see Section~\ref{sec:tl_t_T}.

\subsection{Mixed Erlang distribution}\label{sec:ME}
In our specific case, $f_1$ is the density of a mixed Erlang (ME) distribution which is lower truncated at $t^l\geq0$ and upper truncated at $t>t^l$. 
More specifically, we consider a mixture of $M$ Erlang distributions with common scale parameter $\theta>0$.
\par The Erlang distribution is a Gamma distribution with an integer shape parameter. It has density function and cumulative distribution function
\begin{equation*}
f_E(x ; r, \theta) = 
\frac{x^{r-1} \exp(-x/\theta)}{\theta^{r}(r-1)!} \quad \mbox{ and } \quad F_E(x; r, \theta) = 1 -  \sum_{z=0}^{r-1}  \exp(-x/\theta) \frac{(x/\theta)^{z}}{z!}\\
\end{equation*}
for $x > 0$, where $r$, a positive integer, is the shape parameter, and $\theta>0$ is the scale parameter. Its inverse $\lambda=1/\theta$ is called the rate parameter.

 \par The density of the ME distribution is then given by
\begin{equation*}
f_1^*(x ; \bs{\alpha}, \bs{r}, \theta) = \sum_{j=1}^{M} \alpha_j \frac{x^{r_j-1} \exp(-x/\theta)}{\theta^{r_j}(r_j-1)!} = \sum_{j=1}^{M} \alpha_j f_E(x; r_j, \theta)  \qquad \mbox{for } x > 0 \, ,
\label{eq:mixerlang_pdf}
\end{equation*}
where the positive integers $\bs{r}=(r_1,\ldots,r_M)$ with $r_1<\ldots<r_M$ are the shape parameters of the Erlang distributions, and $\bs{\alpha}=(\alpha_1,\ldots,\alpha_M)$, with $\alpha_j>0$ and $\sum_{j=1}^{M} \alpha_j=1$, are the mixing weights. 
Similarly, the cumulative distribution function can be written, for $x > 0$, as
\[F_1^*(x ; \bs{\alpha}, \bs{r}, \theta) = \sum_{j=1}^{M} \alpha_j \left(1-\sum_{z=0}^{r_j-1} \exp(-x/\theta) \frac{(x/\theta)^{z}}{z!}\right)= \sum_{j=1}^{M} \alpha_j F_E(x; r_j, \theta).
\]
\par After truncation, with limits $t^l$ and $t$, the probability density function becomes
\begin{align*}
 f_1(x ; t^l, t, \bs{r},  \bTheta_1) = \begin{cases} \displaystyle\frac{ f_1^*( x ; \bs{r}, \bTheta_1^*)}{ F_1^*( t  ; \bs{r}, \bTheta_1^*) -  F_1^*( t^l  ; \bs{r}, \bTheta_1^*)} =   \sum_{j=1}^{M} \beta_j   f_E^t(x;  t^l, t,  r_j, \theta) & \quad \text{for } t^l\leq x \leq t\\
0 & \quad \text{otherwise,} \end{cases} \label{eq:truncated_pdf}
\end{align*}
with $\bTheta_1=(\bs{\beta},\theta)$, which is again a mixture with mixing weights 
  \begin{equation}
\beta_j = \alpha_j  \frac{F_E( t  ; r_j, \theta) -  F_E( t^l  ; r_j, \theta)}{ F_1^*( t  ;  \bs{r},\bTheta_1^*) -  F_1^*( t^l  ;  \bs{r},\bTheta_1^*)} \label{eq:beta}
\end{equation}
and component density functions
\begin{equation*}
f_E^t(x;  t^l, t,  r_j, \theta) = \frac{  f_E(x; r_j, \theta)}{F_E( t  ; r_j, \theta) -  F_E( t^l  ; r_j, \theta)}  \, . \label{eq:f_trunc}
\end{equation*}
The component density functions $f_E^t(x;  t^l, t,  r_j, \theta)$ are truncated versions of the original component density functions $f_E(x;  r_j, \theta)$. We obtain the weights $\beta_j $ by reweighting the original weights $\alpha_j$ using the probability of the corresponding mixing component to lie in the truncation interval.
Denote by $F_E^t$ the CDF corresponding to $f_E^t$. The CDF corresponding to $f_1$ is then given by
\begin{equation}\label{eq:ME_trunc_cdf}
F_1(x; t^l, t, \bs{r}, \bTheta_1) = \begin{cases}
0 & \text{if } x \leq t^l \\
 \sum_{j=1}^{M} \beta_j F_E^t(x;t^l,t, r_j, \theta) = \sum_{j=1}^{M} \beta_j  \frac{  F_E(x; r_j, \theta)-  F_E( t^l  ; r_j, \theta)}{F_E( t  ; r_j, \theta) -  F_E( t^l  ; r_j, \theta)}  & \text{if }t^l < x < t \\
1 & \text{if } x \geq t.
\end{cases}
\end{equation}

\par The number of Erlang mixtures $M$ and the positive integer shapes $\bs{r}$ are fixed when estimating $\bTheta_1=(\bs{\beta},\theta)$. 
They are chosen using the approach described in Section~4 of \citet{MME}. A short overview of this approach is included in Section~1.4 in the online addendum.

\subsection{Pareto distribution}\label{sec:Pareto}
The second density $f_2$ is the density of the truncated Pareto distribution with scale parameter $t>0$, shape parameter $\gamma>0$ and upper truncation point $T$ that can be $+\infty$. 
Note that the scale parameter $t$ coincides with the fixed lower truncation point of the tail distribution. As mentioned before, we determine it in advance using an extreme value analysis, see Section~\ref{sec:tl_t_T}. Hence, $\bTheta_2=\gamma$.
More precisely, we have
\[f_2(x;t,T,\gamma) = \frac{f_2^*(x;t,\gamma)}{F_2^*(T;t,\gamma)} = \begin{cases}
\frac{\frac1{\gamma t}\left(\frac{x}{t}\right)^{-\frac1{\gamma}-1}}{1- \left(\frac{T}{t}\right)^{-\frac1{\gamma}}} & \text{if } t <x <T \\
0  & \text{otherwise}, \\
\end{cases}
\]
and
\begin{equation}\label{eq:Pa_cdf}
F_2(x;t,T,\gamma)=\begin{cases}
0 & \text{if } x \leq t \\
\frac{1- \left(\frac{x}{t}\right)^{-\frac1{\gamma}}}{1- \left(\frac{T}{t}\right)^{-\frac1{\gamma}}}  & \text{if } t <x <T \\
1 & \text{if } x \geq T.
\end{cases}
\end{equation}

\section{Fitting a general splicing model to censored and truncated data using the EM algorithm}\label{sec:EMcens}

In this section, we discuss maximum likelihood estimation for fitting a general splicing model, as proposed in Section~\ref{sec:gensplicing}, to censored and/or truncated data. The special case of a splicing model that combines a mixed Erlang distribution (as introduced in Section~\ref{sec:ME}) and a Pareto distribution (Section~\ref{sec:Pareto}) is treated in the subsequent section. The parameters to be estimated are contained in the vector $\bTheta=(\pi,\bTheta_1,\bTheta_2)$. 

\subsection{Randomly censored data}\label{sec:classes}

We represent the censored, and possibly truncated, sample by $\mathcal{X} = \{(l_i,u_i) \cond i=1,\ldots, n\}$, where $l_i$ and $u_i$ denote the lower and upper censoring points of each data point from the sample of size $n$. These censoring points must be interpreted as the lower and upper endpoints of the interval that contains the data point $x_i$, which is not always observed. 
The censoring status of each data point is determined as follows:
\[
	\begin{array}{ll}
			  \text{Uncensored:}						&	 t^l\leq l_i=x_i=u_i\leq T \\
				\text{(Interval) censored:}	&	 t^l \leq l_i<u_i\leq T. \\
	\end{array}
\]
\noindent Left censored and right censored data points can be treated as a special case of interval censored data points with $l_i=t^l$ and $u_i=T$, respectively. In the splicing context, we make a distinction between five cases of data points:

\begin{figure}[H]
\begin{minipage}{0.55\textwidth}
\renewcommand{\theenumi}{\textit{\roman{enumi}}}
\begin{enumerate}[leftmargin=0.6cm]
	\item \label{enum:uncensored}  Uncensored with $t^l \leq l_i=x_i=u_i \leq t<T$
	\item \label{enum:ii}  Uncensored with $t^l<t<l_i=x_i=u_i\leq T$
	\item \label{enum:ME_belowt} Interval censored with $\ithree$
	\item \label{enum:iv}  Interval censored with $\ifour$
	\item \label{enum:overt} Interval censored with $\ifive$.
\end{enumerate}
\end{minipage}\hfill
\begin{minipage}{0.45\textwidth}
	\centering
	\includegraphics[width=\textwidth]{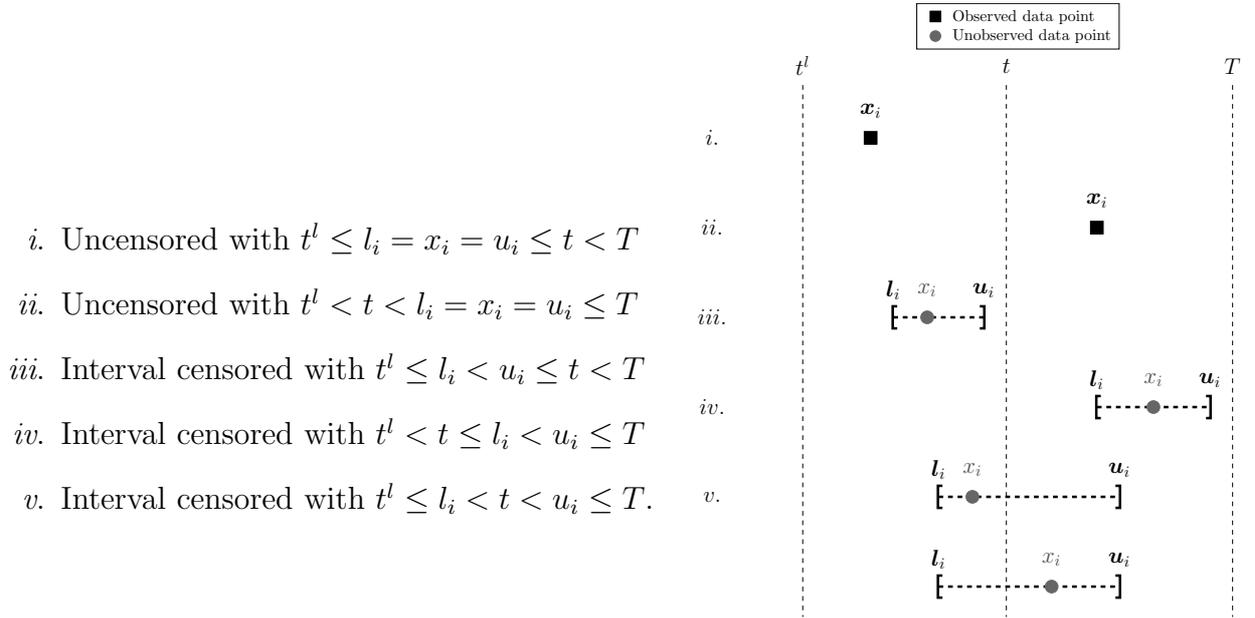}%
	\caption{The different cases of data points.}\label{fig:classes}
\end{minipage}
\end{figure}%
\noindent These cases are visualised in Figure~\ref{fig:classes}. For case \ref{enum:overt}, there are two possible sub-cases: the unobserved data point lies below, or above the splicing point $t$. However, since $x_i$ is not observed then, we cannot distinguish between the two sub-cases. 

\subsection{Maximum likelihood estimation using the EM algorithm}

We use maximum likelihood to fit the parameters of the spliced distribution. 
The likelihood function of the parameter vector $\bTheta$ is given by
\begin{align*}
\mathcal{L} (\bTheta; \mathcal{X}) =& \, \prod_{i \in S_{i.}}  \pi  f_1(x_i; t^l, t, \bTheta_1)\ 
\prod_{i \in S_{ii.}} (1-\pi)  f_2(x_i; t, T, \bTheta_2)  \nonumber \\
& \prod_{i \in S_{iii.}} \pi  \left( F_1(u_i; t^l, t, \bTheta_1) - F_1(l_i; t^l, t, \bTheta_1) \right) \nonumber  \\ 
& \prod_{i \in S_{iv.}} (1-\pi) \left( F_2(u_i; t, T, \bTheta_2) - F_2(l_i; t, T, \bTheta_2) \right)  \nonumber \\ 
& \, \prod_{i \in S_{v.}} \left( \pi + (1-\pi)  F_2(u_i; t, T, \bTheta_2)  -  \pi  F_1(l_i; t^l, t, \bTheta_1) \right), 
\end{align*}%
where $S_{i.}$ is the subset of $\{1,\ldots,n\}$ corresponding to data points of case \ref{enum:uncensored}, and similarly for the other cases.  
The corresponding log-likelihood is
\begin{align}
\ell(\bTheta; \mathcal{X}) =& \, \sum_{i \in S_{i.}}  \left( \ln \pi  + \ln f_1(x_i; t^l, t, \bTheta_1)  \right) + 
\sum_{i \in S_{ii.}} \Big( \ln(1-\pi) +  \ln f_2(x_i; t, T, \bTheta_2)  \Big)   \nonumber \\
+& \sum_{i \in S_{iii.}} \Big( \ln\pi + \ln \left( F_1(u_i; t^l, t, \bTheta_1) - F_1(l_i; t^l, t, \bTheta_1) \right)  \Big)  \nonumber  \\ 
+& \sum_{i \in S_{iv.}} \Big( \ln(1-\pi) + \ln \Big( F_2(u_i; t, T, \bTheta_2) - F_2(l_i; t, T, \bTheta_2) \Big)   \Big)  \nonumber \\ 
+& \, \sum_{i \in S_{v.}}  \ln \Big( \pi + (1-\pi)  F_2(u_i; t, T, \bTheta_2)  -  \pi  F_1(l_i; t^l, t, \bTheta_1) \Big) . 
\label{eq:loglik}
\end{align} 

Direct numerical optimisation of the log-likelihood expression \eqref{eq:loglik} is not straightforward due to the censoring. Data points corresponding to case \ref{enum:overt}, where the censoring interval contains the splicing point $t$, lead to logarithmic terms of a sum involving the splicing weight $\pi$, the parameters $\bTheta_1$ of the body distribution as well as the parameters $\bTheta_2$ of the tail distribution of the splicing model. This prevents separate optimisation with respect to each of these parameter blocks. 

We use the EM algorithm to overcome this hurdle in fitting a splicing model to censored data. 
This iterative method, first introduced by \citet{EM}, finds the maximum likelihood estimates when the data are incomplete and direct likelihood maximisation is not easy to perform numerically. Consider the complete data $\mathcal{Y}$ containing the uncensored sample $\bs{x}=(x_1,\ldots, x_n)$. Given the complete version of the data, we can construct a complete likelihood function as
\begin{align*}
\mathcal{L}_{\text{complete}} (\bTheta;\mathcal{Y})&= \prod_{i=1}^n \left(\pi f_1(x_i; t^l, t, \bTheta_1) \right)^{I(x_i\leq t)}\prod_{i=1}^n \Big((1-\pi) f_2(x_i; t, T, \bTheta_2) \Big)^{I(x_i>t)},
\end{align*}
where $I(x_i\leq t)$ is the indicator function for the event $x_i\leq t$. 
The corresponding complete data log-likelihood function is 
\begin{align}\label{eq:ll_complete} 
\ell_{\text{complete}} (\bTheta;\mathcal{Y}) &=  \sum_{i=1}^n I(x_i\leq t) \left(\ln \pi + \ln f_1(x_i; t^l, t, \bTheta_1) \right) \nonumber\\
&\quad+\sum_{i=1}^n I(x_i>t) \Big(\ln(1-\pi) + \ln f_2(x_i; t, T, \bTheta_2) \Big).
\end{align}
The complete version of the log-likelihood \eqref{eq:ll_complete}, as opposed to the incomplete version \eqref{eq:loglik}, is easy to optimise as it does no longer contain any CDF terms due to censored data points and allows for a separate optimisation with respect to $\pi$, $\bTheta_1$ and $\bTheta_2$. 

However, as we do not fully observe the complete version $\mathcal{Y}$ of the data sample, the complete log-likelihood is a random variable. Therefore, it is not possible to directly
optimise the complete data log-likelihood. The intuitive idea of the EM algorithm for obtaining parameter estimates
in case of incomplete data is to take the conditional expectation of the complete data log-likelihood given the incomplete data and then use this
expected log-likelihood function to estimate the parameters. However, taking the expectation of the complete
data log-likelihood requires the knowledge of the parameter vector, so an iterative approach is needed.

More specifically, starting from an initial guess for the parameter vector, $\bTheta^{(0)}$, the EM algorithm iterates between two steps. In the $h$th iteration of the E-step, we compute the conditional expectation of the complete data log-likelihood with respect to the complete data $\mathcal{Y}$ given the observed data $\mathcal{X}$ and using the current estimate of the parameter vector $\Theta^{(h-1)}$  as true values:
\[E\left(   \ell_{\text{complete}} (\bTheta;\mathcal{Y})    \cond   \mathcal{X} ; \bTheta^{(h-1)}\right).
\]
In the M-step, we maximise the conditional expectation of the complete data log-likelihood obtained in the E-step with respect to the parameter vector:
\[ \bTheta^{(h)} = \operatorname*{arg\,max}_{\bTheta}\ E\left(   \ell_{\text{complete}} (\bTheta;\mathcal{Y})    \cond   \mathcal{X} ; \bTheta^{(h-1)}\right)  \, . \]
Both steps are iterated until convergence. We use a numerical tolerance value of $10^{-3}$ in the data examples in Section~\ref{sec:examples}.

We discuss these steps in detail for a general splicing model in the presence of random censoring in Appendix~\ref{sec:appDetails}.

\section{Fitting the ME-Pareto model}\label{sec:ME_Par}

The fitting procedure for the special case of a splicing model that combines a mixed Erlang distribution and a Pareto distribution is treated in the online addendum. 
The incompleteness now stems on the one hand from censoring and on the other hand from the mixing of Erlang components. The general fitting procedure is therefore extended using ideas from the procedure of \citet{ME} for fitting the ME distribution to interval censored and/or truncated data.
In this section we discuss how the estimation algorithm simplifies in case of no censoring and comment on the selection of splicing and truncation points.

\subsection{Uncensored data}\label{sec:uncens}

When no censoring is present, we only have data points from cases \ref{enum:uncensored} and \ref{enum:ii}. Hence, the EM steps for $\pi$, the ME part and the Pareto part can be performed separately since the parts of the log-likelihood \eqref{eq:loglik} containing $\pi$, $\bTheta_1$ and $\bTheta_2$, respectively, can then be split. We discuss this simplified setting in Section~2 in the online addendum. The splicing weight $\pi$ simply gets estimated as the proportion of data points smaller than or equal to the splicing point $t$.
The algorithm of \citet{ME} is applied to fit a ME distribution to all data points smaller than or equal to $t$. In case there is no upper truncation, i.e.~$T = + \infty$, the solution for $\gamma$  is the Hill estimator \citep{Hill} with threshold $t$. This estimator is commonly used to estimate the shape parameter $\gamma$ when modelling the tail with the Pareto distribution.

\subsection{Selection of splicing and truncation points} \label{sec:tl_t_T}

Up to now, we assumed that the lower truncation point $t^l$, the splicing point $t$ and the upper truncation point $T$ are known. In many applications, there is no lower or upper truncation and we set $t^l = 0$ and $T = + \infty$. 

\par If lower truncation is present, this boundary can often be deduced from the context. For example, in insurance, in case there is a common deductible, the lower truncation point is set to the value of this deductible. 

\par The splicing point $t$ might not always be as straightforward to determine. We do not propose to estimate it using a likelihood approach \citep[see e.g.][]{LN_Pa, Weibull_Pa2, EM_exp_GPD}. Rather, we use extreme value analysis to give an expert opinion about the choice of the splicing point. More specifically, we use the mean excess plot \citep{SoE} to visualise where a transition from the body to the tail of the distribution is suitable, i.e.~to detect different parts of the distribution. This is done by looking for a point $t$ beyond which the mean excess plot is linearly increasing. This then suggests that a Pareto distribution is appropriate to describe the losses $X$ given $X>t$. We demonstrate this type of modelling in the data examples in Section~\ref{sec:examples}. Alternatively, in case of no censoring or upper truncation, adaptive methods based on extreme value theory (EVT) are available for choosing the threshold $t$, see Chapter~4 in \citet{SoE} and references therein. However, such methods might lead to inappropriate threshold choices for splicing models as they do not explicitly try to identify the different distributional parts.  We illustrate this problem with the Danish fire example in Section~\ref{sec:danish}.

\par In situations where the upper truncation point $T$ cannot be set based on the characteristics of the problem, as is for example the case for earthquake magnitudes, we need  a strategy to decide whether upper truncation is applicable to the considered problem, and if so, an estimator for $T$ is required.
\citet{trHill} propose a conditional maximum likelihood estimator (MLE) for the $\gamma$ parameter and the upper truncation point~$T$ of a truncated Pareto distribution when $T$ is unknown. \citet{Truncation} further extend this methodology and provide an improved estimator for $T$. Both papers also suggest a formal test to decide between a truncated and a non-truncated tail distribution.
We explain these methods in Section~2 in the online addendum.
This approach can only be applied in case there is no censoring. For censored data, there is no method available to estimate the parameters of a truncated Pareto distribution when $T$ is unknown.

\section{Risk measures}\label{sec:riskmeasures}

In order to quantify the risk exposure of a company, several risk measures, such as the Value-at-Risk (VaR) and the Tail Value-at-Risk (TVaR), have been developed. Moreover, these risk measures can be used to determine the amount of capital to hold as a buffer against unexpected losses.
\par When estimating the risk measures using statistical methods, it is essential that the fitted model captures the data well. Especially a good fit of the tail part is crucial since this corresponds to the largest losses. A global fit, hence not only a tail fit, is needed as one might be interested in computing reinsurance premiums or performing a risk analysis where focus is not only on extreme events. Further details on the estimation of risk measures can be found in \citet{QRM,LossModels,LossModels_FT,Reins}.

\subsection{Excess-loss insurance premiums}
Using a fitted splicing model such as the ME-Pa model presented in this paper, we calculate premiums for an excess-loss insurance. For this type of insurance, the (re)insurer covers all losses above a certain retention level $R$. This means that she pays $(X-R)_+=\\\max\{X-R,0\}$, where $X$ is the total claim amount. The loss for the insured (also called the cedent) is thus limited to $R$. This type of contract is typical in reinsurance where the reinsurer acts as the insurer's insurer and covers the losses of an insurance company above the retention level. 
The net premium of such an insurance contract is given by
\begin{equation}\label{eq:premium}
\Pi(R;t^l,t,T,\bTheta) = E((X-R)_+)= \int_R^{T} (1-F(z;t^l,t,T,\bTheta))\, dz.
\end{equation}
For $t\leq R<T$ we get
\begin{align*}
\Pi(R;t^l,t,T,\bTheta) &= \int_R^{T} \big(1-(\pi+(1-\pi)F_2(z;t,T,\bTheta_2))\big)\, dz =(1-\pi)\Pi_2(R;t,T,\bTheta_2),
\end{align*}
whereas for $t^l \leq R< t$ we have
\begin{align*}
\Pi(R;t^l,t,T,\bTheta)&= \int_R^t \big(1-\pi F_1(z;t^l,t,\bTheta_1)\big)\, dz + \int_t^{T} \big(1-(\pi+(1-\pi)F_2(z;t,T,\bTheta_2))\big)\, dz \\
&=(t-R) - (t-R)\pi + \pi \int_R^t (1-F_1(z;t^l,t,\bTheta_1))\, dz + (1-\pi)\Pi_2(t;t,T,\bTheta_2)\\
& =(1-\pi) (t-R) + \pi \Pi_1(R;t^l,t,\bTheta_1) + (1-\pi)\Pi_2(t;t,T,\bTheta_2).
\end{align*}
Note that $\Pi(R;t^l,t,T,\bTheta)=\Pi(t^l;t^l,t,T,\bTheta)+(t^l-R)$ for $R< t^l$ and $\Pi(R;t^l,t,T,\bTheta)=0$ for $R\geq T$.
\par We can rewrite
\begin{align*}
\Pi_1(R;t^l,t,\bTheta_1) &= \int_R^t \left(1- \frac{F_1^*(z;\bTheta_1)-F_1^*(t^l;\bTheta_1)}{F_1^*(t;\bTheta_1)-F_1^*(t^l;\bTheta_1)} \right)\, dz \\
&=\frac{F_1^*(t;\bTheta_1)(t-R) -\int_R^t F_1^*(z;\bTheta_1)\, dz}{F_1^*(t;\bTheta_1)-F_1^*(t^l;\bTheta_1)}\\
&=\frac{\left(F_1^*(t;\bTheta_1)-1\right)(t-R) + (\Pi_1^*(R;\bTheta_1)-\Pi_1^*(t;\bTheta_1))}{F_1^*(t;\bTheta_1)-F_1^*(t^l;\bTheta_1)}
\end{align*}
for $t^l\leq R<t$. For the ME distribution, the premium is given by
\[\Pi_1^*(R;\bs{\alpha},\theta)=\theta^2\sum_{m=1}^M\left(\sum_{l=m}^{M-1}\left(\sum_{j=l+1}^M \alpha_j\right)\right)f_E(R;m,\theta)\]
for $R\geq 0$, see \citet{ME}. They assume, without loss of generality, that $r_m=m$ for $m=1,\ldots, M$. Note that $\Pi_1(R;t^l,t,\bTheta_1)=\Pi_1(t^l;t^l,t,\bTheta_1)+(t^l-R)$ for $R < t^l$ and $\Pi_1(R;t^l,t,\bTheta_1)=0$ for $R\geq t$.
\par Similarly, we get
\begin{align*}
\Pi_2(R;t,T,\bTheta_2) = \frac{\left(F_2^*(T;\bTheta_2)-1\right)(T-R)  + (\Pi_2^*(R;\bTheta_2)-\Pi_2^*(T;\bTheta_2))}{F_2^*(T;\bTheta_2)-F_2^*(t;\bTheta_2)}
\end{align*}
for $t\leq R<T$. For the Pareto distribution we have the following premium when $R \geq t$:
\[\Pi_2^*(R;t,\gamma)=\int_R^{+\infty} \left(\frac{z}{t}\right)^{-\frac1{\gamma}}\, dz = R^{-\frac1{\gamma}+1}\frac{t^{\frac1{\gamma}}}{\frac1{\gamma}-1}.\] 
Note that $\Pi_2(R;t,T,\bTheta_2)=\Pi_2(t;t,T,\bTheta_2)+(t-R)$ for $R < t$ and \\$\Pi_2(R;t,T,\bTheta_2)=0$ for $R\geq T$.

\subsection{VaR and TVaR}
The Value-at-Risk (VaR) is a popular risk measure and is defined as a quantile of the distribution, $\VaR_{1-p}= F^{-1}(1-p)$. For the spliced distribution, the quantile function is
\[F^{-1}(p;t^l,t,T,\bTheta) = \begin{cases}
F_1^{-1}(p/\pi;t^l,t,\bTheta_1) & \quad \text{if } 0 \leq p \leq \pi \\
F_2^{-1}\big((p-\pi)/(1-\pi);t,T,\bTheta_2\big)  & \quad \text{if } \pi < p \leq 1.
\end{cases}
\]
The quantile function of the ME distribution $F_1^{-1}$ cannot be computed analytically, but can be obtained by numerically inverting the CDF. For the (truncated) Pareto distribution we have
\begin{align*}
F_2^{-1}(p;t,T,\gamma) = F_2^{*\, -1}\big(p F_2^*(T;t,\gamma);t,\gamma\big)=t \left(1-p+p\left(\frac{T}{t}\right)^{-\frac1\gamma}\right)^{-\gamma}.
\end{align*}

\par Closely related is the Tail Value-at-Risk (TVaR) which is defined as the expected loss given that the loss is larger than $\VaR_{1-p}$. When the CDF is continuous in $\VaR_{1-p}$, which is the case for our spliced CDF since it is continuous everywhere, the TVaR can be rewritten as \citep[see e.g.][]{LossModels}
\begin{align*}
\TVaR_{1-p} := E\left(X \cond X> \VaR_{1-p}\right) =\VaR_{1-p} + \frac{\Pi(\VaR_{1-p})}{p}.
\end{align*}
This can thus easily be computed using the expressions for $\VaR_{1-p}$ and $\Pi(R)$.

\section{Data examples}\label{sec:examples}

\subsection{Danish fire insurance data}\label{sec:danish}

Our first data example concerns the Danish fire insurance dataset \citep{Danish} from the Copenhagen Reinsurance Company which contains information on 2167 fire losses from 1980 to 1990. It can be found in the \texttt{R} package \textit{evir} \citep{evir}. The claim sizes are expressed in millions of Danish kroner (DKK) and are adjusted for inflation to reflect values in 1985. Only claims that are larger than 1 million kroner are included. This means that left truncation occurs at 1. This dataset has already been considered by several other authors including \citet{McNeil_EVT, Embrechts, EM_exp_GPD}.
Some authors, among other \citet{LN_Pa, LN_Pa2, Weibull_Pa2, Mixtures_IME}, consider additional observations below 1 and set the left truncation point at 0.  Rather than using specific ad hoc combinations of (mixtures of) standard light-tailed distributions with a heavy-tailed distribution, or ad hoc mixtures of distributions,  we here illustrate the generic ME-Pareto splicing method and compare this global fit with the result of a pure ME fit. This global fit is then used to provide an estimate for the premium of an excess-loss insurance with a certain retention $R$.
\begin{figure}[ht]
		\centering
		\includegraphics[height=0.6\textwidth,angle=270]{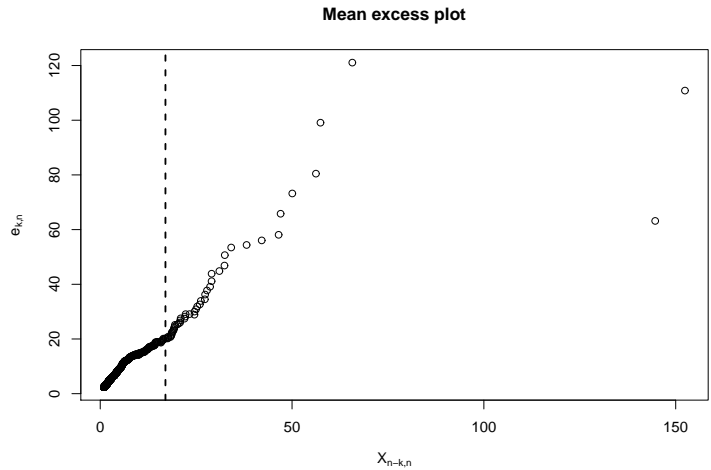}%
		\caption{Danish fire insurance: Mean excess plot.}\label{fig:danish_ME}
	\end{figure}
	
\par The splicing point $t$ is chosen based on the mean excess plot \citep{SoE}. This plot consists of estimates for the mean excess values
\begin{equation}\label{eq:meanexcess}
e(v)=E\left(X-v \cond X>v\right) = \frac{\int_v^{+\infty} (1-F(x))\, dx}{1-F(v)},
\end{equation} in the order statistics $v=X_{n-k,n}=\hat{Q}\left(1-\frac{k+1}{n+1}\right)=\hat{Q}\left(\frac{n-k}{n+1}\right)$ with $k=1,\ldots, n-1$, where the CDF $F$ is estimated by the empirical CDF  $\hat{F}$, and $\hat{Q}$ is the corresponding empirical quantile function. 
The linear increasing parts in Figure~\ref{fig:danish_ME} suggest a Pareto tail although the last two observations are behaving differently. The splicing point is chosen at $t = 17$,
as indicated by the vertical dashed line, since the mean excess slope changes at this point. A Pareto distribution is suitable to model the loss distribution after $t=17$ since the mean excess plot is linearly increasing from this point on. Therefore, this is a suitable point for the transition from the body to the tail of the distribution.
	
\par We fit the ME-Pareto splicing model starting from $M=10$, and consider spread factors $s\in\{1,\ldots,10\}$ (see \citet{MME} and Section~1.1 in the online addendum).
The full fitting procedure took 6.89s using R~3.4.0 \citep{R} on Windows 7 (64-bit) OS with an Intel Core i7-3770 CPU @ 3.40GHz.
The fitted model was obtained using $s=10$ and is summarised in Table~\ref{tab:param_danish}. It consists of a mixture of three Erlang distributions for the body and the Pareto distribution for the tail. The estimates for $\bs \beta$ corresponding to $\hat{\bs \alpha}$ are equal to $\hat{\bs \beta}=(0.819, 0.152, 0.029)$.
	\begin{table}[H]
	\centering
	\begin{tabular}{c|c|c}
	\hline
	Splicing & ME & Pareto \\ \hline
	$\begin{aligned}[t]
	\hat{\pi}&=0.976\\
	t^l&=\num{1} \\
	t&=\num{17} \\
	T&=+\infty\\
	 \end{aligned}$ & 
		$\begin{aligned}[t]
		\hat{\bs \alpha}&=(0.938,0.051, 0.011)\\
		\hat{\bs r}&=(1,6,16)\\
		\hat{\theta}&=\num{0.811}\\
		\\
		\end{aligned}$ & $\hat{\gamma}=0.530$ \\\hline
	\end{tabular}
		\caption{Danish fire insurance: summary of the fitted ME-Pa splicing model.}\label{tab:param_danish}
	\end{table}

\par
\begin{figure}[ht]
		\makebox[\linewidth][c]{%
		\begin{subfigure}{0.5\linewidth}
			\centering
			\includegraphics[height=\textwidth,angle=270]{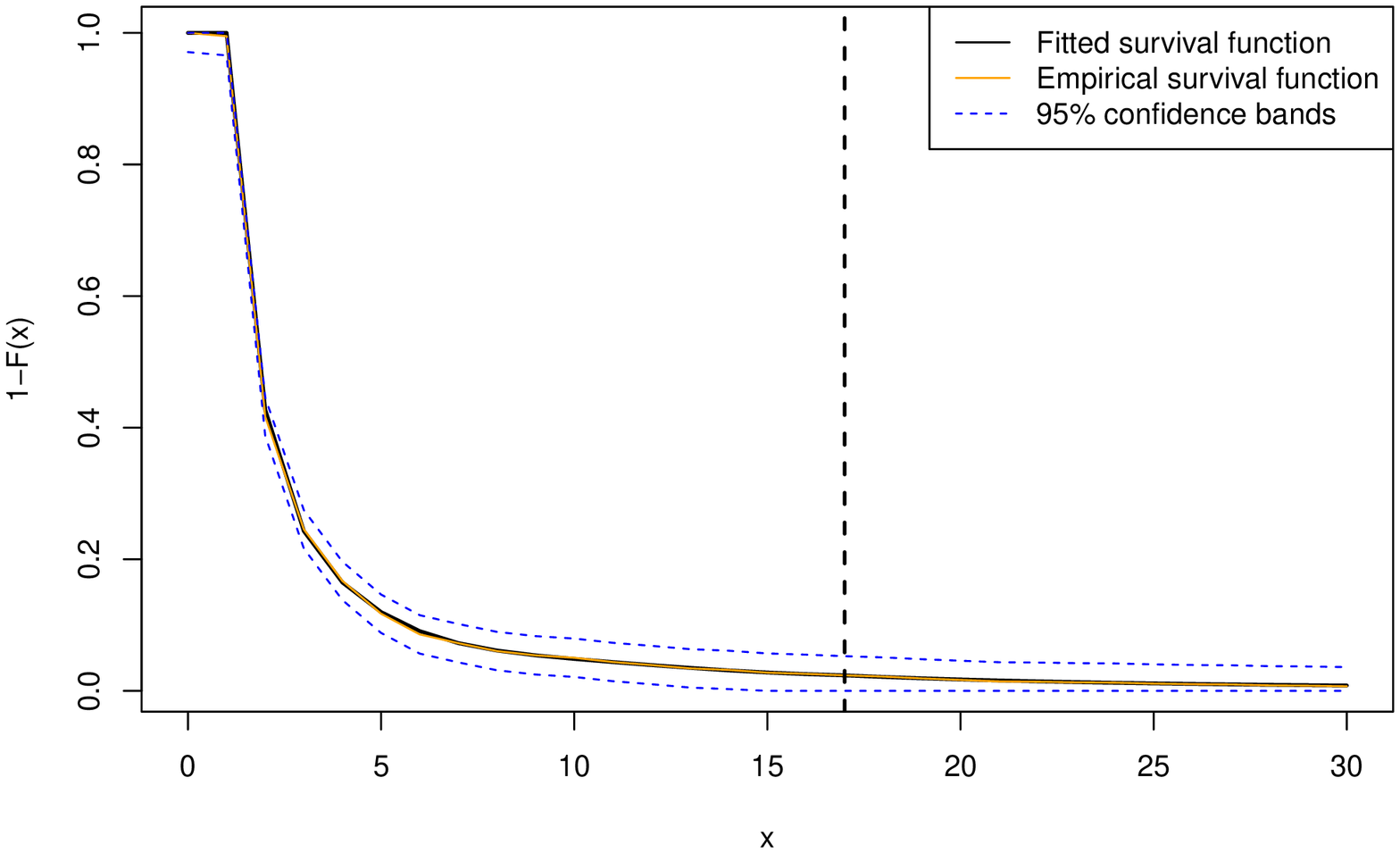}%
			\caption{}\label{fig:danish_surv}
		\end{subfigure}
		~
		\begin{subfigure}{0.5\linewidth}
			 \includegraphics[height=\textwidth,angle=270]{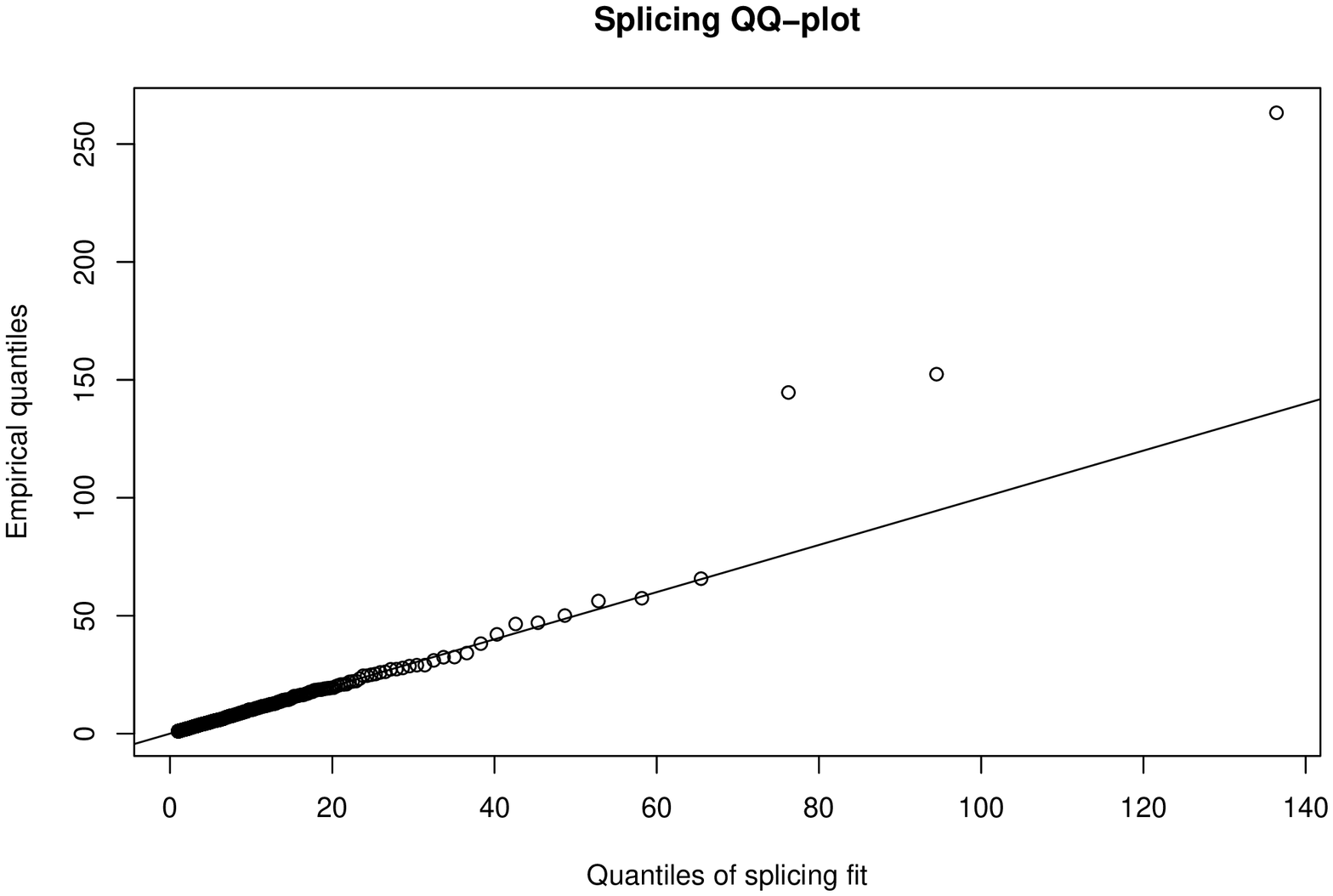}%
			 \caption{}\label{fig:danish_QQ}
		\end{subfigure}}%
		\caption{Danish fire insurance: (a) Survival plot  and (b) QQ-plot of the fitted ME-Pa splicing model.}%
	\end{figure}

\begin{figure}[ht]
		\makebox[\linewidth][c]{%
		\begin{subfigure}{0.5\linewidth}
			\centering
			\includegraphics[height=\textwidth,angle=270]{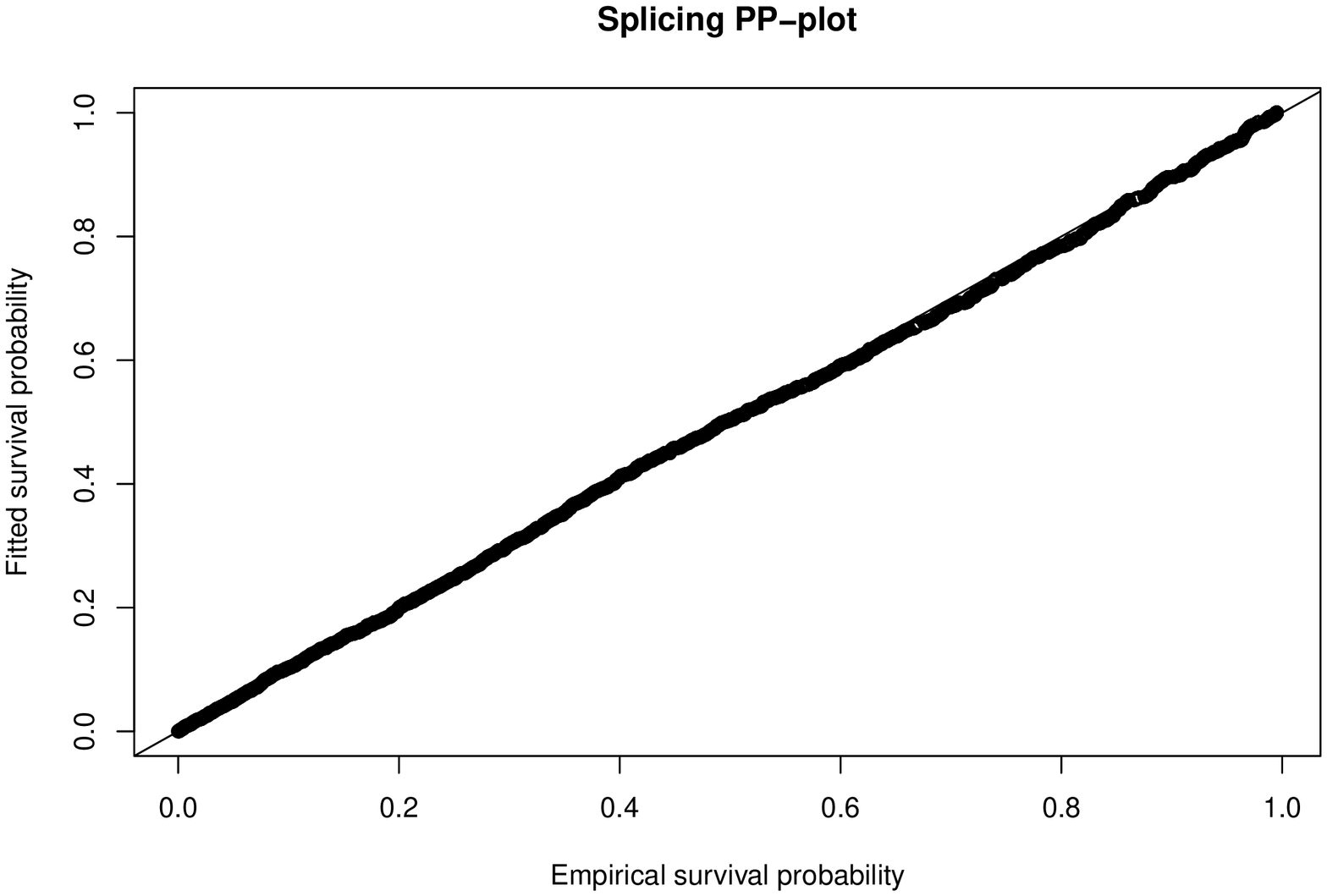}%
			\caption{}\label{fig:danish_PP}
		\end{subfigure}
		~
		\begin{subfigure}{0.5\linewidth}
			\centering
			\includegraphics[height=\textwidth,angle=270]{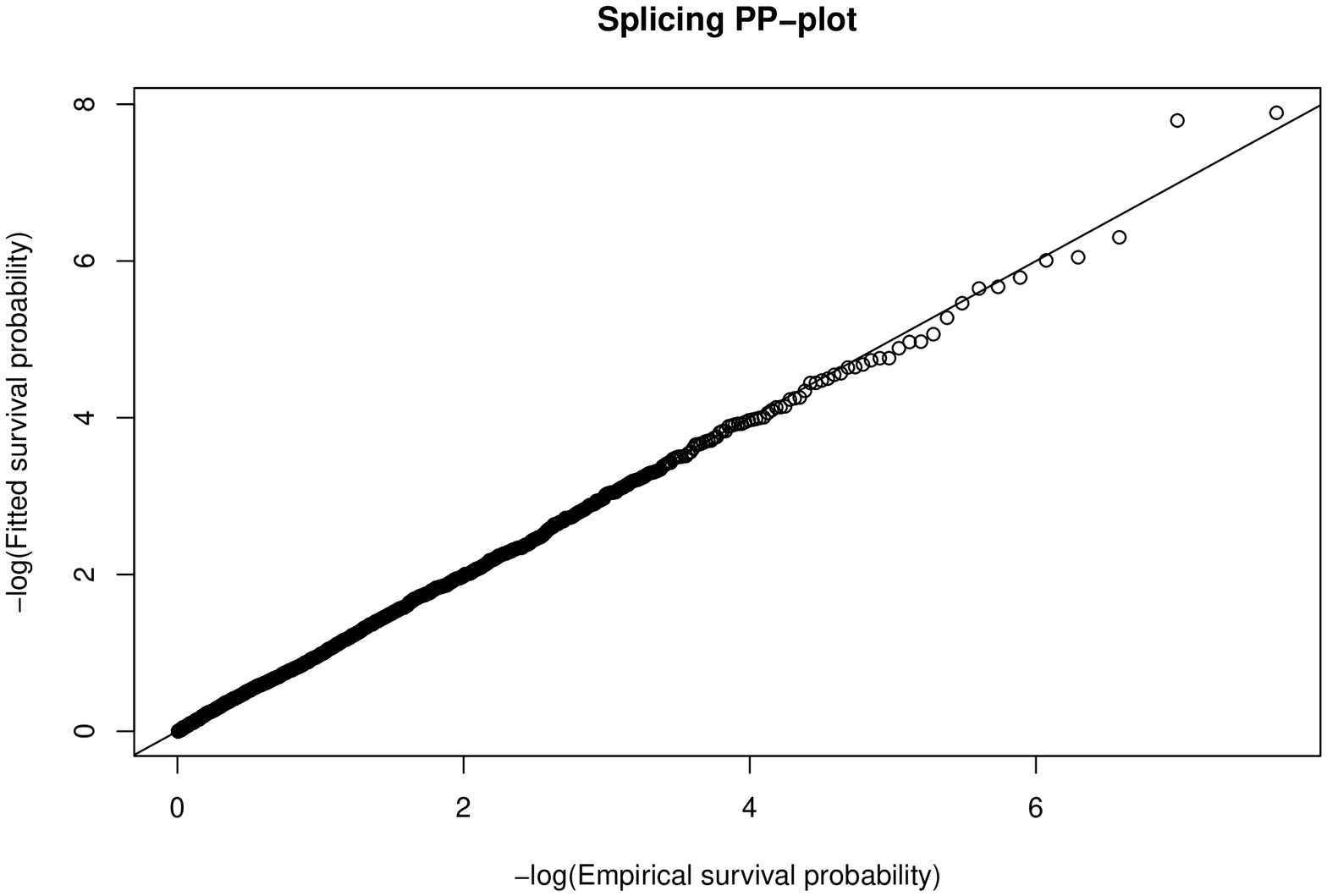}%
			\caption{}\label{fig:danish_PP_log}
		\end{subfigure}}%
		\caption{Danish fire insurance: PP-plots of the fitted  ME-Pa splicing model with (a) ordinary and (b) minus-log scale.}%
	\end{figure}

 In order to evaluate the splicing fit with the ME and Pareto distributions, graphical tools, information criteria and goodness-of-fit (GoF) tests are considered.
A first graphical tool is the survival plot in Figure~\ref{fig:danish_surv} where the fitted survival function (black) is plotted together with the empirical survival function (orange). 95\% confidence bands for the empirical estimator (dashed blue) and a vertical line indicating the splicing point are also added. These confidence bands are determined using the Dvoretzky-Kiefer-Wolfowitz inequality \citep{DKW}. The fitted spliced survival function follows the empirical survival function closely and lies well within the confidence bands. Next, to inspect this fit in more detail, a QQ-plot is constructed (Figure~\ref{fig:danish_QQ}) comparing the empirical quantiles to the fitted quantiles. All but the last three points on the QQ-plot are close to the 45 degree line suggesting a good fit. Fitting these three points is challenging as noted e.g.~by \citet{EM_exp_GPD}. Closely related is the PP-plot in Figure~\ref{fig:danish_PP} where the fitted survival function is plotted vs.~the empirical survival function. This plot confirms that the model gives a good global fit. However, it is difficult to asses the quality of the tail fit from the PP-plot. Therefore, a PP-plot with a minus-log scale is also constructed (Figure~\ref{fig:danish_PP_log}). The upper right corner then corresponds to the tail of the distribution. As expected, there are some deviations from the 45 degree line for the largest points, but the plot still indicates a good global fit.

Additional to the graphical tools, we look at the negative log-likelihood (NLL), AIC and BIC values for each model where lower values are better, see Table~\ref{tab:danish_GoF}. The AIC and BIC values are defined as
\[\mbox{AIC}= 2\times\mbox{NLL} + 2\times df \quad \text{and} \quad \mbox{BIC}= 2\times \mbox{NLL} + \ln n\times df\]
where $df$ denotes the degrees of freedom, i.e.~the number of estimated parameters in the model.
Moreover, we consider the Kolmogorov-Smirnov (KS) and Anderson-Darling (AD) GoF statistics as they are a measure for the distance between the empirical CDF and the fitted CDF of a model. The KS statistic is defined as
\[D_n=\mbox{sup}_{x\geq t^l}\,|\hat{F}_n(x)-F(x)|\]
where $\hat{F}_{n}$ is the empirical CDF based on $n$ observations and $F$ the fitted CDF.
The AD statistic is given by
\[A_n=n \int_{t^l}^{+\infty} \frac{(\hat{F}_{n}(x)-F(x))^2}{F(x)(1-F(x))}\, dx.\]
Note that both test statistics take lower truncation at $t^l$ into account.
These statistics are commonly used to test if the data sample is drawn from a specified (continuous) distribution. The standard P-values of the test are not valid when the model parameters are estimated from the data \citep{GoF}. Therefore, we use a bootstrap approach that is detailed in \citet{GoF} and \citet{LossModels}. First, we compute the KS and AD test statistics using the fitted model for the Danish fire insurance data. Then, we generate 1000 samples with replacement from the Danish fire insurance data. For each sample, the model is fitted and then the KS and AD statistics are computed. The P-values are then obtained as the proportion of these 1000 test statistics that exceed the test statistic computed in the first step.
The \texttt{R} \citep{R} packages \texttt{stats} (KS) and  \texttt{ADGofTest} \citep{ADGofTest} (AD) are used to compute the test statistics. The results are also displayed in Table~\ref{tab:danish_GoF} where values closer to 0 indicate a better fit. The corresponding P-values are added between brackets.
Apart from the fitted splicing model, we also consider the following models:
\begin{itemize}
	\item The ME and GPD splicing model (ME-GPD) with the same splicing point as before: $t=17$. Hence, it has the same ME distribution for the body of the distribution as the ME-Pa model. The tail of the distribution is modelled by a GPD with parameters $\hat{\gamma}=0.654$ (shape) and $\hat{\sigma}= 7.917$ (scale).
		\item The ME model with $t^l=1$, $\bs\alpha=(0.8807, 0.0878, 0.0165, 0.0087, 0.0039, 0.0017, 0.0005, \\0.0002)$, $\bs r=(2,   8,  20,  33,  52,  94, 269, 477)$ and $\theta=0.553$.
\end{itemize}
Note that the ME-GPD model is fitted using our general fitting procedure. The ME fit is obtained using the approach in \citet{ME} starting from $M=25$ and considering spread factors $s\in\{1,\ldots,20\}$.
\par The NLL and AIC values in Table \ref{tab:danish_GoF} of the ME fit are lower than those of the ME-Pa and ME-GPD fits. However, the ME is overfitting the data as many of the eight Erlang components have a small weight and describe only one or a few observations in the tail. Because MEs have asymptotically exponential tails, there is no parsimonious model possible using such mixtures to provide an appropriate fit for this heavy-tailed data. Erlang components in a mixture are not able to extrapolate the heaviness in the tail and instead behave similar to an empirical distribution in the upper tail, which is undesirable from a risk measurement perspective. This behaviour of ME is also illustrated using the simulated GPD sample in \citet{ME} and motivates our approach. BIC, which penalises the number of components more than AIC, indicates that the splicing models have a better trade-off between the number of parameters and the quality of the fit. The ME-Pa and ME-GPD models do not have this drawback as the Pareto distribution or GPD are appropriate for tail extrapolation based on extreme value theory. To avoid the overfitting problem, less components (e.g.~3) can be used, but the ME model is then not able to provide an appropriate fit for the heavy-tailed data.
The values of the NLL suggest that the ME-GPD fit is slightly better than the ME-Pa fit. However, when taking both the quality of the fit and the number of parameters into account, as is done in AIC and BIC, the ME-Pa model is preferred over the ME-GPD model. The P-values of the GoF tests are large for all models suggesting that all models provide an appropriate fit for the data. Based on the graphical tools, the ICs, and the P-values of the KS and AD tests, we propose to use the ME-Pa model when modelling the Danish fire insurance data. 

\begin{table}[ht]
\centering
\begin{tabular}{r|ccc|cc}
  	\hline
	Model & NLL & AIC & BIC & KS & AD \\ 
	\hline 
	ME-Pa & \num{3327.332}  &  \num{6670.663} & \textbf{\num{6716.112}} & 0.025 (0.643) & 1.424 (0.558)  \\
	ME-GPD & \num{3327.122} & \num{6672.244}  & \num{6723.374} & 	0.025 (0.643) & 1.423 (0.555) \\
	ME & \textbf{\num{3317.702}} & \textbf{\num{6667.405}} & \num{6758.302} &  0.015 (0.956) & 0.405 (0.926) \\
   \hline
\end{tabular}
\caption{Danish fire insurance: NLL, AIC and BIC values, and GoF test statistics and P-values.}\label{tab:danish_GoF}
\end{table}

\par Next, we also compare the results with the lognormal-Pareto (LN-Pa) and Weibull-Pareto (W-Pa) fits as introduced by \citet{LN_Pa2,Weibull_Pa2}, respectively, where the threshold is chosen adaptively in the likelihood procedure. Applying these methods for choosing $t$ in our setting leads to a threshold choice around 1.8. The mean excess plot in Figure~\ref{fig:danish_ME} shows different distributional parts, but this choice for $t$ does not properly take these into account, leading to an inferior global fit. While the NLL, AIC and BIC values (NLL=\num{3331.063}, AIC= \num{6668.127} and BIC=\num{6685.17} for LN-Pa, and NLL=\num{3330.774}, AIC=\num{6667.548} and BIC=\num{6684.591} for W-Pa) are comparable with the values obtained from the ME-Pa splicing fit, the QQ-plots of the fitted LN-Pa and W-Pa models in the online addendum show that the fit is less appropriate for values above 10. Moreover, using an adaptive selection method based on the asymptotic mean squared error of the Hill estimator, see Section~4.7 in \citet{SoE}, leads to a choice of $t=1.38$. The ME-Pa fit using this splicing point has the same problem as the discussed LN-Pa and W-Pa fits showing that this threshold choice also does not take the different distributional parts into account.

\par As an illustration, premiums for excess-loss insurances can be computed using the fitted models. Table~\ref{tab:danish_premium} shows the computed premiums for different models and different retentions. Additional to the three previously mentioned models, premiums are also computed non-parametrically using \eqref{eq:premium} with the empirical survival function (Non-par.), and using the combination of a non-parametric fit for the body (below $t=17$ as before) and the Pareto distribution for the tail (Non-par.--Pa).  All three parametric models result in premiums that are close to the ones obtained using the non-parametric model when the retention levels are small. For higher levels the estimates are substantially different. The non-parametric model results in zero premiums when the retention levels are larger than the maximal data value, 263.2504. Due to its exponential tail, the ME distribution results in lower premium estimates for high retentions than the heavy-tailed ME-Pa and ME-GPD models. The Non-par.--Pa and ME-Pa models have the same fit for the tail, but a different fit for the body. Therefore, the premium estimates for high retentions are the same, but the premiums for retentions below the splicing point $t=17$ differ. Although the ME-Pa and ME-GPD models have the same model for the body of the distribution, the estimates for the premiums also differ for low retentions since the survival function is integrated starting from the retention level when estimating the premiums, see \eqref{eq:premium}.
\begin{table}[ht]
\centering
\begin{tabular}{rrrrrr}
  \hline
R & Non-par. & Non-par.--Pa & ME-Pa & ME-GPD & ME \\ 
  \hline
1 & 2.3851 & 2.3657 & 2.3657 & 2.4531 & 2.3851 \\ 
  5 & 1.0630 & 1.0436 & 1.0485 & 1.1359 & 1.0597 \\ 
  10 & 0.7083 & 0.6889 & 0.6884 & 0.7757 & 0.7080 \\ 
  50 & 0.2029 & 0.1727 & 0.1727 & 0.2678 & 0.2003 \\ 
  100 & 0.1201 & 0.0933 & 0.0933 & 0.1803 & 0.1206 \\ 
	200 & 0.0292 & 0.0504 & 0.0504 & 0.1232 & 0.0295 \\ 
  300 & 0 & 0.0352 & 0.0352 & 0.0989 & 0 \\ 
	\hline
\end{tabular}
\caption{Danish fire insurance: estimates for premiums of excess-loss insurance with different retentions $R$ (expressed in millions of Danish kroner).}\label{tab:danish_premium}
\end{table}

\subsection{Motor third party liability insurance}
The second data example consists of motor third party liability (MTPL) insurance claims in Europe between 1995 and 2010 \citep{Reins}.
They are evaluated at the end of 2010, i.e.~right before the beginning of 2011, and 59\% of the 837 claims are not closed at that time. 
All amounts are indexed in order to reflect costs in calendar year 2011, with inflation taken into account. Our goal is again to provide a good overall fit and to estimate excess-loss insurance premiums.

\par
\begin{figure}[ht]
		\centering
		\includegraphics[height=0.5\textwidth]{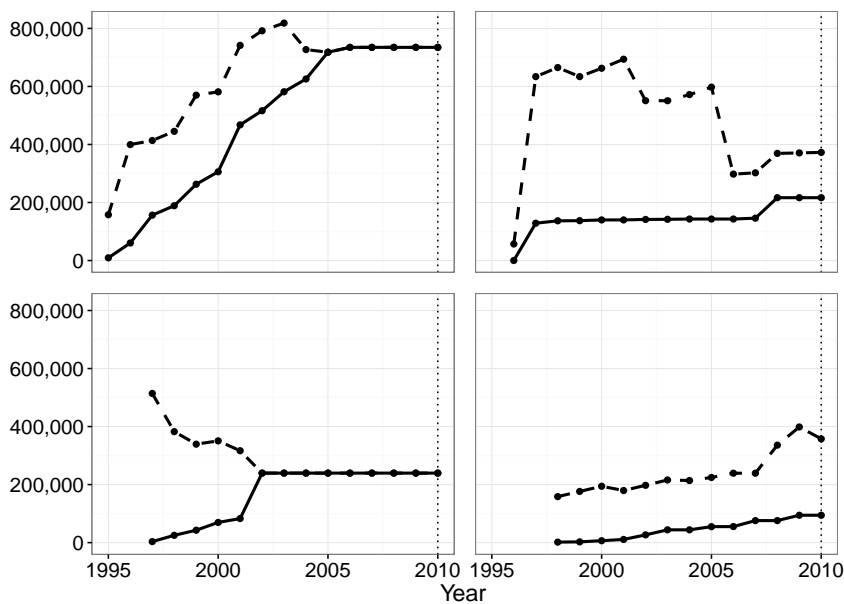}%
		\caption{MTPL: cumulative indexed payments (full line) and indexed incurred values (dashed line) at the end of each year for four claims. The moment of evaluation, i.e.~the end of 2010, is indicated by the vertical dashed line.}\label{fig:MTPL_paths}
	\end{figure}

 As discussed in Section~\ref{sec:intro}, a significant time may elapse between the claim occurrence and its final settlement (due to e.g.~legal procedures or severe bodily injury). In order to illustrate the development of a claim, in Figure~\ref{fig:MTPL_paths}, we show for four claims the cumulative indexed payment (full line) and the indexed incurred (dashed line) at the end of each year. The incurred at the end of a given year is equal to the sum of the cumulative payment up to that moment and an expert's estimate for the outstanding loss. The claims occurred, respectively, in 1995, 1996, 1997 and 1998. The first and third claim are closed before the end of the observation period (indicated by the vertical dashed line in Figure~\ref{fig:MTPL_paths}), and hence the cumulative indexed payment and the indexed incurred value at the end of 2010 are equal.
The second and fourth claim are still in development at the end of 2010 and the indexed incurred is larger than the cumulative indexed payment at that moment.

We apply the splicing approach for censored data using an interval censoring framework with the cumulative indexed payments at the end of 2010 as lower bound for the final cumulative indexed payment. It makes sense to construct an upper bound based on the incurreds since they are determined conservatively using information on the specific claim: e.g.~the severity of the accident, the number of people involved. As an illustration of the method, and by lack of further claim information, we use here the indexed incurreds at the end of 2010  as upper bound. However, when a claim is early in development, i.e.~there is a small period between the claim occurrence and the moment of evaluation, the incurreds might still be too uncertain to be used as an upper bound since the information available to the expert might be limited.  After several years of development, the quality of the incurreds has improved a lot, as more information becomes available, making them more suitable as an upper bound. Since claims with accident years between 2006 and 2010 are still early in development, and we do not have more information to improve their incurreds, we omit them for the analysis \citep[as is done in][]{Reins}. We then have 596 claims left and 45\% of them are not closed at the end of 2010. A more prudent approach is to only use the cumulative indexed payments at the end of 2010 as lower bound in a right censoring framework. However, this does not take the valuable information of incurreds into account. Another possibility is to ignore any censoring information and to only consider the indexed incurreds when estimating the final claim amount. In this example we compare these three possible strategies.
\par
\begin{figure}[ht]
\makebox[\linewidth][c]{%
		\begin{subfigure}{0.5\linewidth}
			\centering
			\includegraphics[height=\textwidth,angle=270]{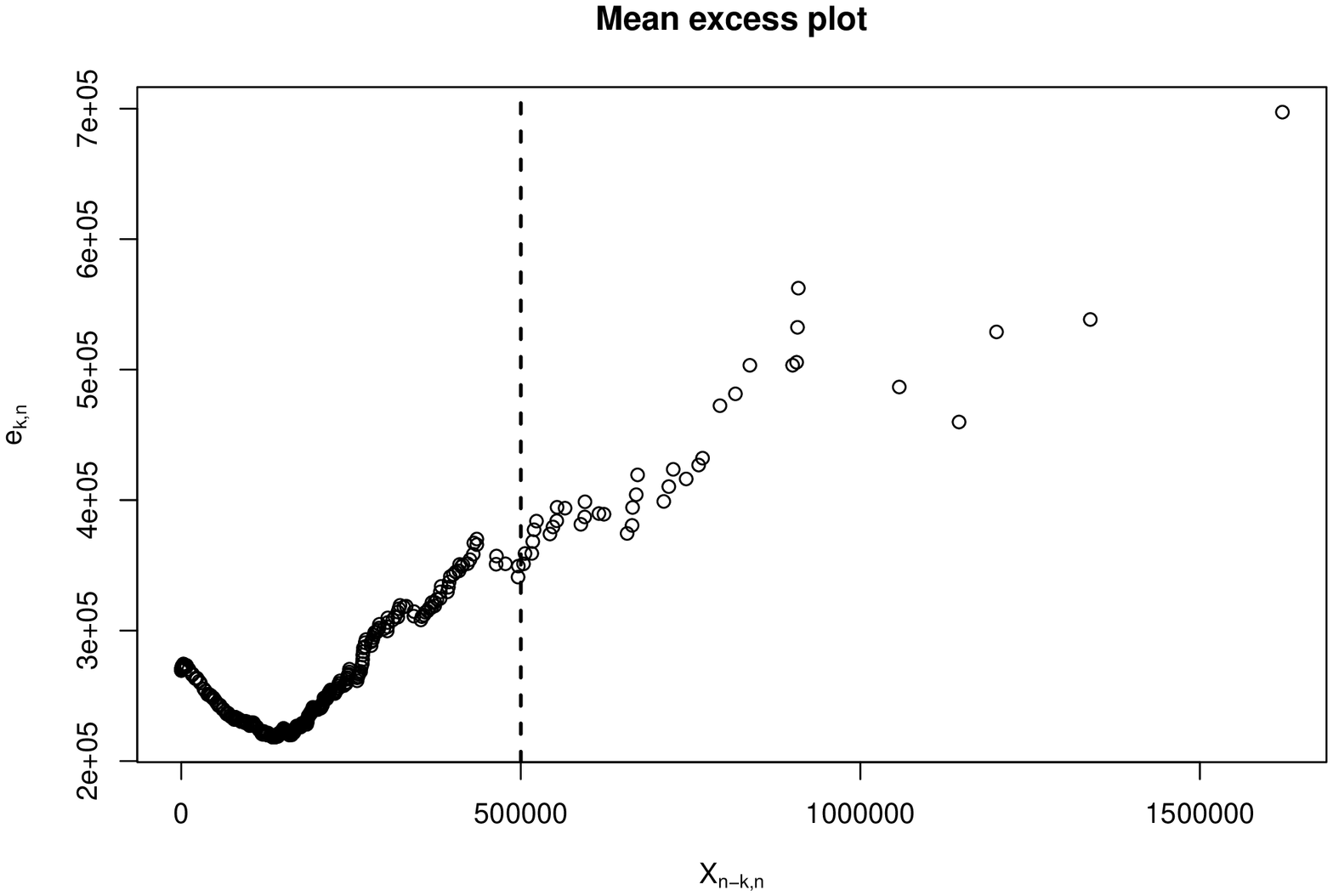}%
			\caption{}\label{fig:MTPL_ME}
		\end{subfigure}
		~
		\begin{subfigure}{0.5\linewidth}
			 \centering
			 \includegraphics[height=\textwidth,angle=270]{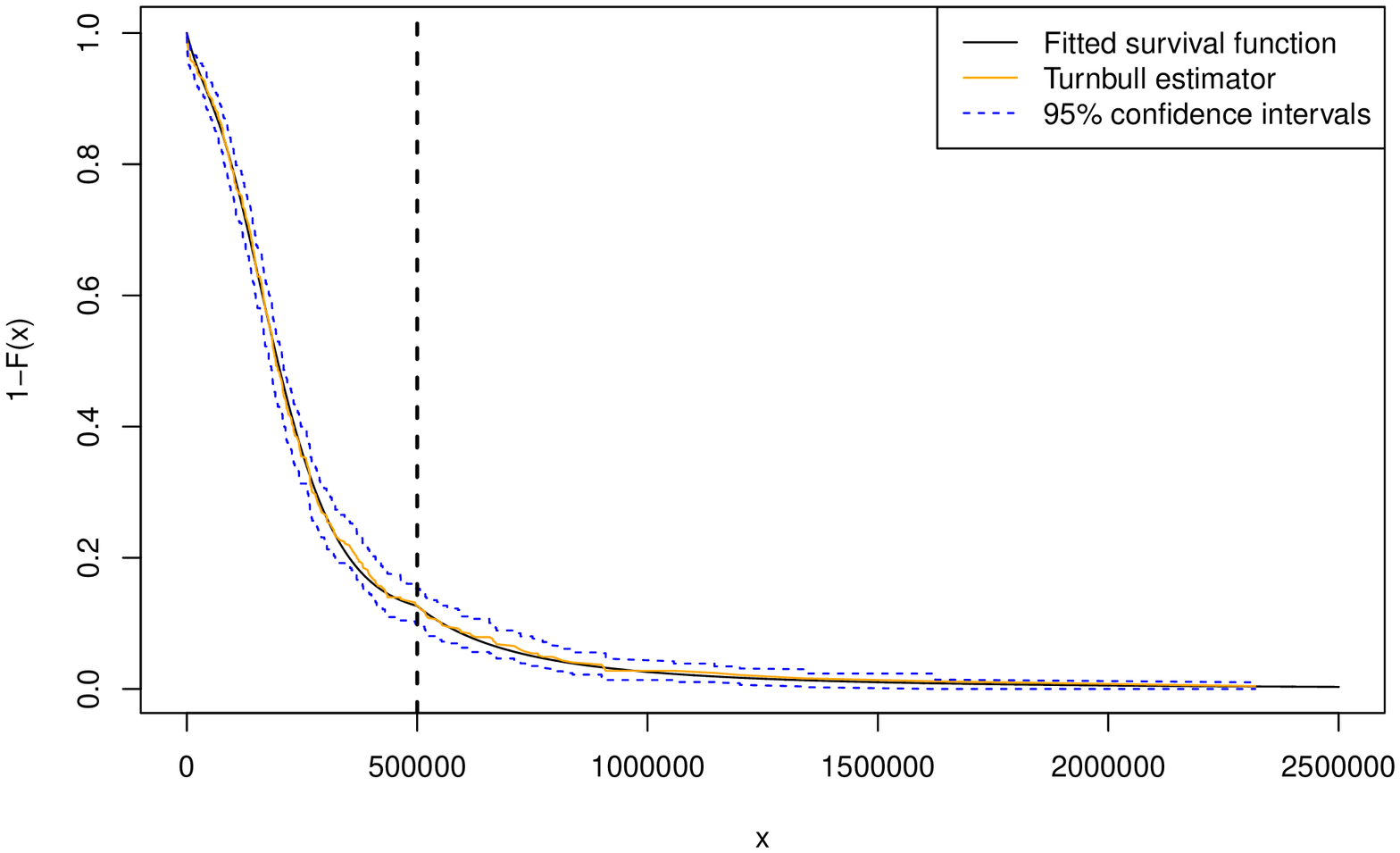}%
			 \caption{}\label{fig:MTPL_surv}
		\end{subfigure}	}%
		\caption{MTPL: (a) Mean excess plot based on the Turnbull estimator and (b) survival plot of the fitted ME-Pa splicing model.}%
	\end{figure}

 As before, we rely on the mean excess plot to choose the splicing point $t$. We now use the Turnbull estimator \citep{Turnbull76} to estimate the distribution function in \eqref{eq:meanexcess}.
This is a non-parametric estimator for the CDF in the case of interval censored data points. It extends the Kaplan-Meier estimator \citep{KM}, which can only be used for right censored data, to interval censored data. There is no analytical solution for the Turnbull estimator and its computation relies on the EM algorithm. We use the implementation in the \texttt{R} package \texttt{interval}~\citep{interval}.
The resulting mean excess estimates are
\[\hat{e}(v) = \frac{\int_v^{+\infty} (1-\hat{F}^{TB}(x))\, dx}{1-\hat{F}^{TB}(v)},\]
where $\hat{F}^{TB}$ is the Turnbull estimator for the CDF. We evaluate this function in $v=\hat{Q}^{TB}(1-(k+1)/(n+1))=\hat{Q}^{TB}((n-k)/(n+1))$, for $k=1, \ldots, n-1$, where $\hat{Q}^{TB}$ is the estimator for the quantile function based on the Turnbull estimator, since in the uncensored case we also used the empirical quantiles corresponding to $1/(n+1), \ldots, (n-1)/(n+1)$. The estimates are plotted in Figure~\ref{fig:MTPL_ME}. The mean excess plot now has a convex shape indicating that a Pareto tail is suitable. A different slope is visible after \num{500000} and we therefore choose the splicing point at $t=\num{500000}$ as shown by the vertical line. As discussed in Section~\ref{sec:classes}, there are five classes of data points when fitting a splicing model to censored data. Using the splicing point $t=\num{500000}$, the number of data points per class is $\#S_{i.}=296$, $\#S_{ii.}=34$, $\#S_{iii.}=175$, $\#S_{iv.}=25$ and $\#S_{v.}=66$, where $\#S$ denotes the number of data points in a set $S$.
\par The model is fitted starting from $M=10$ and with $s\in\{1,\ldots,10\}$ (see Section~1.1 in the online addendum). The full fitting procedure took 29.31s.
The fitted model consists of $M=2$ Erlangs and was obtained using $s=2$. It is summarised in Table~\ref{tab:param_mtpl}. The estimates for the weights $\bs \beta$ of the truncated ME are in this case equal to $\hat{\bs \beta}=(0.174, 0.826)$.

\begin{table}[ht]
	\centering
	\begin{tabular}{c|c|c}
	\hline
	Splicing & ME & Pareto \\ \hline
	$\begin{aligned}[t]
	\hat{\pi}&=0.873\\
	t^l&=0\\
	t&=\num{500000} \\
	T&=+\infty\\
	 \end{aligned}$ & 
		$\begin{aligned}[t]
		\hat{\bs \alpha}&=(0.171, 0.829)\\
		\hat{\bs r}&= (1,4)\\
		\hat{\theta}&=\num{55227}\\
		\end{aligned}$ & $\hat{\gamma}=0.438$ \\\hline
	\end{tabular}
		\caption{MTPL: summary of the fitted ME-Pa splicing model.}
	\label{tab:param_mtpl}
	\end{table}

\par
\begin{figure}[ht]
	\makebox[\linewidth][c]{%
		\begin{subfigure}{0.5\linewidth}
			\centering
			\includegraphics[height=\textwidth,angle=270]{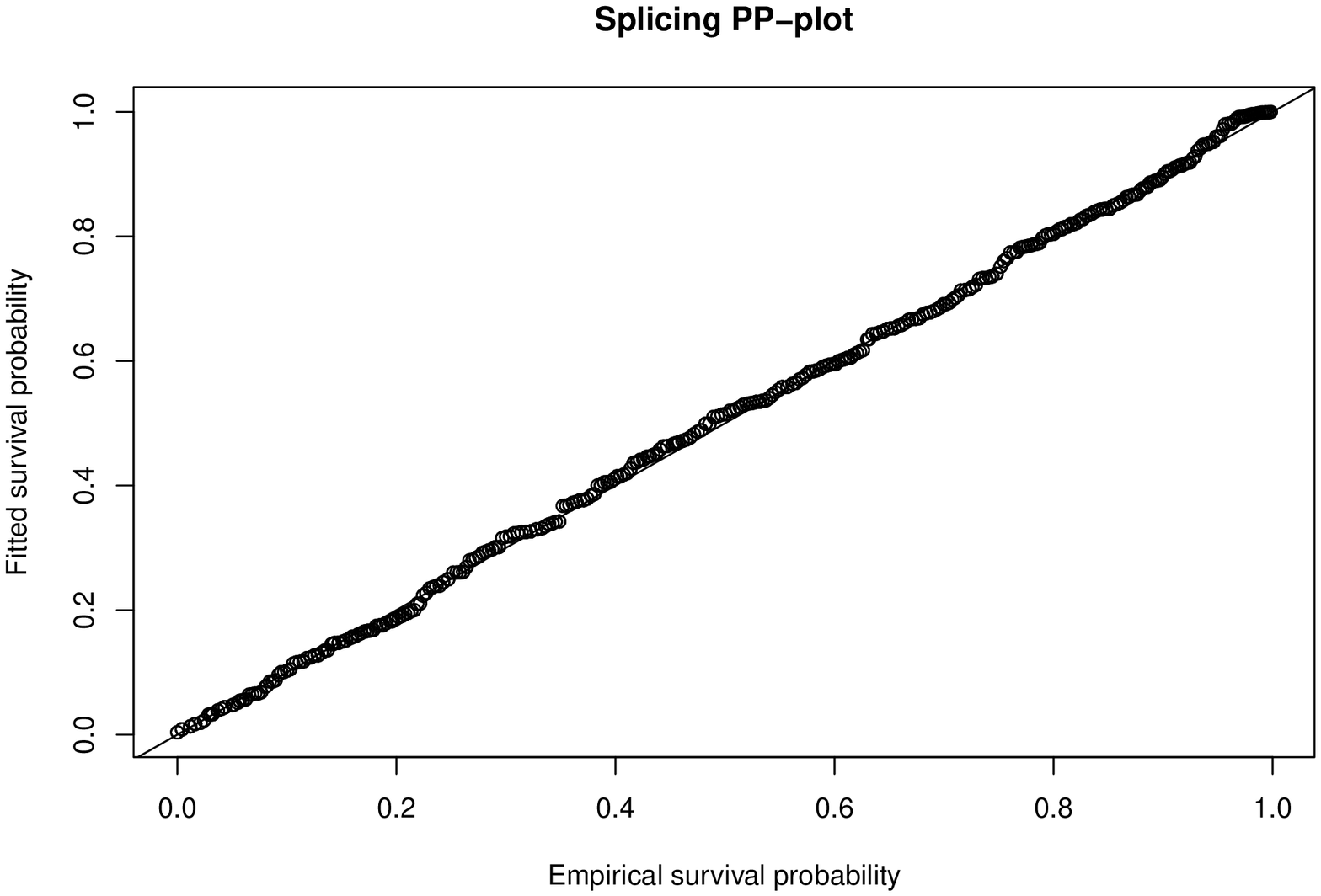}%
			\caption{}\label{fig:MTPL_PP}
		\end{subfigure}
		~
		\begin{subfigure}{0.5\linewidth}
			\centering
			\includegraphics[height=\textwidth,angle=270]{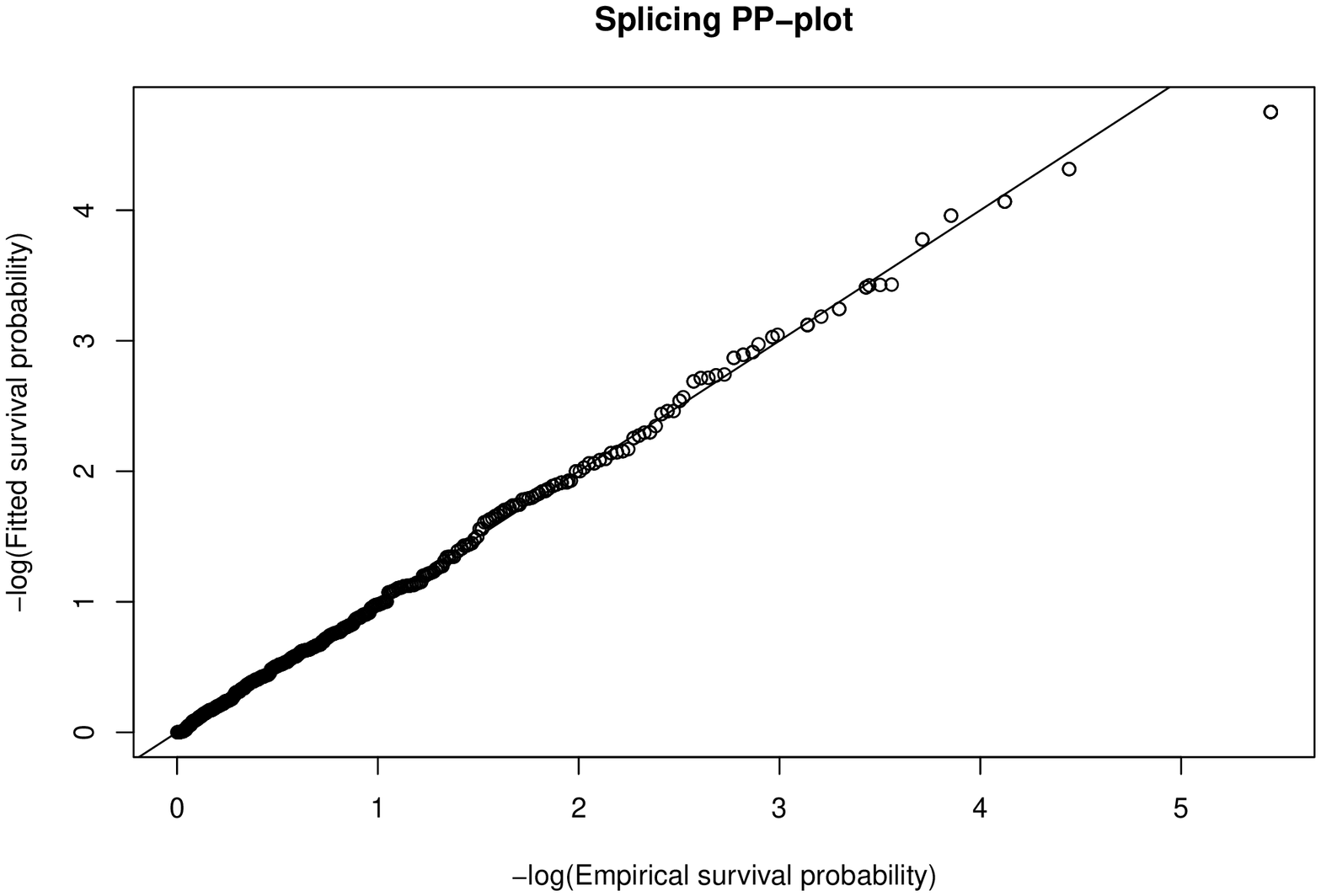}%
		 \caption{}\label{fig:MTPL_PP_log}
		\end{subfigure}}%
		\caption{MTPL: PP-plots of the fitted ME-Pa splicing model with (a) ordinary and (b) minus-log scale.}\label{fig:MTPL_PP_all}
	\end{figure}

 Some of the graphical tools used in Section~\ref{sec:danish} can be extended to the censoring case. The fitted survival function can be compared to the non-parametric Turnbull estimate (Figure~\ref{fig:MTPL_surv}). Pointwise confidence intervals are obtained using 200 bootstrap samples generated by the \texttt{R} package \texttt{interval}~\citep{interval}. They are added as dashed blue lines in Figure~\ref{fig:MTPL_surv}. The fitted survival function follows the Turnbull estimate closely and stays within the confidence intervals suggesting a good fit. PP-plots are made using the fitted survival function and the Turnbull survival function, see Figures~\ref{fig:MTPL_PP} and \ref{fig:MTPL_PP_log}, where a minus-log scale is used in the second plot. 
Both lines are close to the 45 degree line indicating that the fitted model is suitable for the data.
	
\par
\begin{figure}[ht]
		\centering
		\includegraphics[height=0.5\textwidth,angle=270]{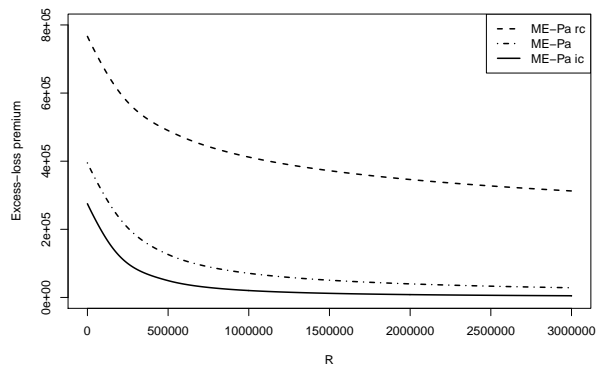}%
		\caption{MTPL: Estimates for premiums of excess-loss insurance with different retentions using ME-Pa splicing model with interval censoring (full line), right censoring (dashed line) and no censoring (dash-dot line).}\label{fig:MTPL_XL}
	\end{figure}

 Figure~\ref{fig:MTPL_XL} shows estimates for premiums of excess-loss insurance for different retentions.
The premiums are estimated using the considered splicing model in the interval censoring framework (full line), and compared to estimates obtained using a splicing model based on the right censoring framework (dashed line) and a splicing model without censoring using the indexed incurreds (dash-dot line). The second approach gives higher premium estimates than the first approach since the (censored) total amount paid for each claim is not bounded from above. The incurreds are conservative expert estimates of the final cumulative claim amount. Only using the indexed incurreds in an uncensored framework does not take into account that the actual total amount that needs to be paid can be lower than the indexed incurreds. Therefore, it leads to higher premium estimates than for the splicing model using interval censored data. Using all information available, the cumulative indexed payments and the indexed incurreds, leads to significantly lower premium estimates.
\par As mentioned in the introduction, a non-parametric fit for the body and a parametric model (e.g.~Pareto distribution) for large losses can be used instead of a splicing model. When censoring is present, this approach can no longer be used as we might have data points of class \ref{enum:overt} (see Figure~\ref{fig:classes}) where the lower bound of the interval is in the body of the distribution, whereas the upper bound is in the tail. As is shown in this example, our general splicing framework can handle observations of this type and can hence be used to provide a global fit. This global fit is e.g.~needed to compute premiums for excess-loss insurances.
\par We discussed the possibility to use the GPD instead of the Pareto distribution in the splicing model. Unlike for the Pareto distribution, the fourth and sixth expectation in the E-step \eqref{eq:Estep} cannot be computed analytically when using the GPD (with $\gamma\neq0$). This makes the whole procedure numerically more intensive as it requires numerical integration. Without censoring, this drawback is not present as only the first two expectations in the E-step \eqref{eq:Estep} need to be computed.

\section{Conclusions}

\par In order to get a suitable global fit for financial loss data we propose a new splicing model. It combines the flexibility of the mixed Erlang distribution to model the body of the distribution with the Pareto distribution to provide a suitable fit for the tail. Hence, our proposal avoids ad hoc combinations of a standard light-tailed distribution for the body with a heavy-tailed distribution for the tail.
\par Motivated by real life insurance data sets where censoring and truncation are omnipresent, we provide a general framework for fitting a spliced distribution to censored and/or truncated data. This fitting procedure uses the EM algorithm to handle data incompleteness due to censoring. Moreover, we give details on the application of this procedure to fit the ME-Pa model.
\par Estimates for excess-loss premiums and risk measures such as the VaR can be easily extended to the splicing context. We illustrate the flexibility of the proposed ME-Pa splicing approach using the lower truncated Danish fire insurance dataset and using the MTPL dataset where censoring is present.
\par As we provide a general procedure to fit a splicing model to censored and/or truncated data, other distributions for the body and/or tail can be considered. We illustrated the use of the GPD instead of the Pareto distribution for the tail.
\par We implemented all models and tools in the \texttt{R} package \texttt{ReIns} which complements \citet{Reins}. The package is available on CRAN: \url{https://CRAN.R-project.org/package=ReIns}.

\section*{Acknowledgements}
The authors are grateful to the referees for their helpful comments and suggestions. The authors thank Hansj\"org Albrecher for the interesting discussions, and the company for providing the data for the MTPL example.
	\par Roel Verbelen acknowledges support from the agency for Innovation by Science and Technology (IWT). Jan Beirlant acknowledges support through a restart project from KU Leuven's research council. Katrien Antonio acknowledges financial support from the Ageas Continental Europe Research Chair at KU Leuven and from KU Leuven's research council [project COMPACT C24/15/001].
\addcontentsline{toc}{section}{Acknowledgements}

\bibliography{bib_splicing}

\addcontentsline{toc}{section}{References}

\appendix

\section{Details on fitting procedure for censored and truncated data}\label{sec:appDetails}

\subsection{Initial step}
Before iterating the EM-steps, we need starting values for the splicing weight $\pi$ and for the parameters of the distributions for the body and the tail: $\bTheta^{(0)}=(\pi^{(0)},\bTheta_1^{(0)},\bTheta_2^{(0)})$. Suitable starting values depend on the distributions used for the body and the tail. We discuss starting values for the splicing of the ME and Pareto distributions in Section~1.1 in the online addendum.

\subsection{E-step}\label{sec:Estep}
In the $h$th iteration of the E-step, we take the conditional expectation of the complete log-likelihood \eqref{eq:ll_complete} given the incomplete data $\mathcal{X}$, the points $t^l$, $t$ and $T$, and the current estimate $\bTheta^{(h-1)}$ for $\bTheta$. We distinguish the five cases of data points again to determine the contribution of a data point to the conditional expectation $E\left(\ell_{\text{complete}} (\bTheta; \mathcal{Y})\cond \mathcal{X}, t^l, t, T; \bTheta^{(h-1)}\right)$:
\renewcommand{\theenumi}{\textit{\roman{enumi}}}
\begin{enumerate}
	\item $\ln \pi+ E\left(\ln f_1(X_i;t^l,t,\bTheta_1) \cond \ione ;\bTheta_1^{(h-1)}\right)$
	\item $\ln(1-\pi) + E\left(\ln f_2(X_i;t,T,\bTheta_2) \cond \itwo ;\bTheta_2^{(h-1)}\right)$
	\item $\ln \pi + E\left(\ln f_1(X_i;t^l,t,\bTheta_1) \cond \ithree ;\bTheta_1^{(h-1)}\right)$
	\item $\ln(1-\pi) + E\left(\ln f_2(X_i;t,T,\bTheta_2) \cond \ifour ;\bTheta_2^{(h-1)}\right)$
	\item $\!
\begin{aligned}[t]
	&E\left(\vphantom{\bTheta^{(h-1)}}\left[\ln \pi + \ln f_1(X_i;t^l,t,\bTheta_1) \right]\, I(\{X_i\leq t\})\right.\\
	&+\left.\left[\ln(1-\pi)+ \ln f_2(X_i;t,T,\bTheta_2) \right]\, I(\{X_i>t\})
	  \cond \ifive ;\bTheta^{(h-1)}\right)
\end{aligned}
$ 
\end{enumerate}
Note that the event $\{\ione\}$ indicates that we know $t^l$, $l_i=u_i$, $t$ and $T$, and that the ordering $t^l\leq l_i=u_i\leq t<T$ holds. Similar reasonings hold for the other conditional arguments in the expectations.
Using the law of total expectation we can rewrite the expectation in \ref{enum:overt}. as 
\begin{align*}
&E\left(\ln \pi + \ln f_1(X_i;t^l,t,\bTheta_1)\cond  \ifivelower;\bTheta_1^{(h-1)}\right) \\
&\qquad\qquad \times P\left(X_i\leq t\cond  \ifive;\bTheta^{(h-1)}\right)\\
&+E\left(\ln(1-\pi) + \ln f_2(X_i;t,T,\bTheta_2) \cond  \ifiveupper;\bTheta_2^{(h-1)}\right) \\
& \qquad \qquad \times P\left(X_i>t \cond  \ifive;\bTheta^{(h-1)}\right),
\end{align*}
 where $\{\ifivelower\}$ denotes that $t^l$, $l_i$, $t$, $u_i$ and $T$ are known, that the ordering $\ifive$ holds, and that $\{X_i \leq t\}$.
The considered conditional expectation of the complete log-likelihood is then given by
\begin{align}\label{eq:Estep}
&E\left(\ell_{\text{complete}} (\bTheta; \mathcal{Y})\cond \mathcal{X}, t^l, t, T; \bTheta^{(h-1)}\right) \nonumber\\
&= \ \sum_{i \in S_{i.}}\Big[\ln \pi+ E\left(\ln f_1(X_i;t^l,t,\bTheta_1) \cond \ione;\bTheta_1^{(h-1)}\right)\Big] \nonumber\\
& \ +\sum_{i \in S_{ii.}}\Big[\ln(1-\pi) + E\left(\ln f_2(X_i;t,T,\bTheta_2) \cond \itwo;\bTheta_2^{(h-1)}\right)\Big] \nonumber\\
&  \ +\sum_{i \in S_{iii.}}\Big[\ln \pi + E\left(\ln f_1(X_i;t^l,t,\bTheta_1) \cond \ithree;\bTheta_1^{(h-1)}\right)\Big]\nonumber\\
&  \ +\sum_{i \in S_{iv.}}\Big[\ln(1-\pi) + E\left(\ln f_2(X_i;t,T,\bTheta_2) \cond\ifour;\bTheta_2^{(h-1)}\right)\Big] \nonumber\\
&  \ +\,\sum_{i \in S_{v.}}\Big[\ln \pi + E\left(\ln f_1(X_i;t^l,t,\bTheta_1)\cond  \ifivelower;\bTheta_1^{(h-1)}\right)\Big]\nonumber\\
&\qquad \qquad  \qquad \qquad \qquad \qquad  \times P\left(X_i\leq t\cond  \ifive;\bTheta^{(h-1)}\right)\nonumber\\ 
&  \ +\,\sum_{i \in S_{v.}} \Big[\ln(1-\pi) + E\left(\ln f_2(X_i;t,T,\bTheta_2) \cond  \ifiveupper ;\bTheta_2^{(h-1)}\right)\Big]\nonumber\\
& \qquad \qquad \qquad \qquad \qquad \qquad \times  P\left(X_i>t \cond  \ifive;\bTheta^{(h-1)}\right).
\end{align}

\noindent Using \eqref{eq:spliceCDF}, the probability in the second to last term in \eqref{eq:Estep} can be written as
\begin{align}
&P\left(X_i\leq t \cond\ifive;\bTheta^{(h-1)}\right) \nonumber\\
&\qquad= \frac{F\left(t;t^l,t,T,\bTheta^{(h-1)}\right)-F\left(l_i;t^l,t,T,\bTheta^{(h-1)}\right)}{F\left(u_i;t^l,t,T,\bTheta^{(h-1)}\right)-F\left(l_i;t^l,t,T,\bTheta^{(h-1)}\right)} \nonumber \\
&\qquad=\frac{\pi^{(h-1)}-\pi^{(h-1)} F_1\left(l_i;t^l,t,\bTheta_1^{(h-1)}\right)}{\pi^{(h-1)}+(1-\pi^{(h-1)})F_2\left(u_i;t,T,\bTheta_2^{(h-1)}\right)-\pi^{(h-1)} F_1\left(l_i;t^l,t,\bTheta_1^{(h-1)}\right)},
\label{eq:prob1}
\end{align}
and the probability in the last term of \eqref{eq:Estep} is given by 1 minus this expression.

\subsection{M-step}

We maximise \eqref{eq:Estep} with respect to $\pi$, $\bTheta_1$ and $\bTheta_2$ by computing the partial derivatives and equating them to zero. In case it is not possible to find analytical solutions for one of these parameters, we need to rely on numerical procedures.

\subsubsection{Maximisation w.r.t.~\texorpdfstring{$\pi$}{pi}}

We use the notations $n_1$ and $n_2$ for the number of data points $X_i$ smaller than or equal to $t$, and above $t$, respectively. 
The partial derivative of \eqref{eq:Estep} w.r.t.~$\pi$ is given by
\[
\frac{\partial E\left(\ell_{\text{complete}} (\bTheta; \mathcal{Y})\cond \mathcal{X}, t^l, t, T, \bTheta^{(h-1)}\right) }{\partial \pi} = \frac{n_1^{(h)}}{\pi} - \frac{n_2^{(h)}}{1-\pi}
\]
with 
\[
n_1^{(h)}=\# S_{i.}+ \#S_{iii.}+\sum_{i \in S_{v.}} P\left(X_i\leq t\cond  \ifive;\bTheta^{(h-1)}\right),
\]
and
\[
n_2^{(h)}=\# S_{ii.}+ \#S_{iv.} + \sum_{i \in S_{v.}} P\left(X_i>t \cond \ifive;\bTheta^{(h-1)}\right).
\]
Data points belonging to case~\ref{enum:overt} are weighted using probabilities \eqref{eq:prob1} and $1-\eqref{eq:prob1}$, leading to 
the estimates $n_1^{(h)}$ and $n_2^{(h)}$ in the $h$th iteration. Note that $n_1^{(h)}+n_2^{(h)}=n.$ Setting the derivative equal to 0 and then solving for $\pi$ yields
\begin{equation}\label{eq:pi_Mstep}
\pi^{(h)}=\frac{n_1^{(h)}}{n_1^{(h)}+n_2^{(h)}}=\frac{n_1^{(h)}}n.
\end{equation}
This updated splicing weight can be interpreted as the proportion of data points smaller than or equal to $t$ as estimated in the $h$th iteration. 

\subsubsection{Maximisation w.r.t.~\texorpdfstring{$\bTheta_1$}{Theta1}}
In order to maximise \eqref{eq:Estep} w.r.t.~$\bTheta_1$, we have to maximise
\begin{align*}
 &\ \sum_{i \in S_{i.}} E\left(\ln f_1(X_i;t^l,t,\bTheta_1) \cond \ione;\bTheta_1^{(h-1)}\right) \nonumber\\
+&\sum_{i \in S_{iii.}} E\left(\ln f_1(X_i;t^l,t,\bTheta_1) \cond \ithree;\bTheta_1^{(h-1)}\right) \nonumber\\
+&\,\sum_{i \in S_{v.}} E\left(\ln f_1(X_i;t^l,t,\bTheta_1)\cond \ifivelower;\bTheta_1^{(h-1)}\right) \nonumber\\
&\qquad \quad\times P\left(X_i\leq t\cond \ifive;\bTheta^{(h-1)}\right).
\end{align*}

\subsubsection{Maximisation w.r.t.~\texorpdfstring{$\bTheta_2$}{Theta2}}
Similarly, to maximise \eqref{eq:Estep} w.r.t.~$\bTheta_2$, we have to maximise
\begin{align*}\label{eq:max_theta2}
&\sum_{i \in S_{ii.}} E\left(\ln f_2(X_i;t,T,\bTheta_2) \cond \itwo;\bTheta_2^{(h-1)}\right) \nonumber\\
+&\sum_{i \in S_{iv.}} E\left(\ln f_2(X_i;t,T,\bTheta_2) \cond \ifour;\bTheta_2^{(h-1)}\right) \nonumber\\
+&\,\sum_{i \in S_{v.}} E\left(\ln f_2(X_i;t,T,\bTheta_2) \cond \ifiveupper ;\bTheta_2^{(h-1)}\right) \nonumber\\
& \qquad\quad\times P\left(X_i>t \cond \ifive;\bTheta^{(h-1)}\right).
\end{align*}

\end{document}


\maketitle

\noindent In this addendum, we give details on application of the framework described in Section~3 applied to the splicing of the ME and Pareto distributions.

\section{Fitting the ME-Pareto model to censored data using the EM algorithm}\label{sec:appEM}

\subsection{Initial step}\label{sec:app_initial}
\par We estimate the splicing constant $\pi$ initially by the Turnbull estimator \citep{Turnbull76} in $t$, i.e.~$\pi^{(0)}=\hat{F}_{TB}(t)$ since the estimate for $\pi$ in the $h$th iteration ($\pi^{(h)}$) can be seen as the proportion of data points smaller than or equal to $t$, see (11).
\par Initial values for the ME parameters follow from \citet{MME} which are improvements of the starting values from \citet{ME}. Consider the data $\bs{d}$ consisting of $x_i$ for all uncensored data points, of $l_i$ for all right censored data points, of $u_i$ for all left censored data points and of the interval midpoints, $\frac{l_i+u_i}2$, for all interval censored data points. Restricting the data $\bs{d}$ to all data points that are smaller than or equal to $t$ gives $\widetilde{\bs{d}}$ which has length $n_{\widetilde{\bs{d}}}$. 
\citet{MME} use the following starting values: $\theta^{(0)}= \frac{\max(\widetilde{\bs{d}})}{s}$, for a given spread factor $s$, 
\begin{equation}\label{eq:rinit}
\bs{r}=\left(\left\lceil\frac{\hat{Q}(0;\widetilde{\bs{d}})}{\theta^{(0)}}\right\rceil,\left\lceil\frac{\hat{Q}\left(\frac1{M-1};\widetilde{\bs{d}}\right)}{\theta^{(0)}}\right\rceil,\ldots,\left\lceil\frac{\hat{Q}(1;\widetilde{\bs{d}})}{\theta^{(0)}}\right\rceil\right)
\end{equation}
 and 
\[\alpha_j^{(0)}=\displaystyle\frac{\sum_{i=1}^{n_{\widetilde{\bs{d}}}} I\left(\left\{ r_{j-1} \theta^{(0)} < \widetilde{d}_i \leq r_j \theta^{(0)}\right\} \right)}{n_{\widetilde{\bs{d}}}}\]
 for $j=1,\ldots,M$, where $r_0=0$ for notational convenience and $\hat{Q}(\cdot;\widetilde{\bs{d}})$ is the empirical quantile function of the data $\widetilde{\bs{d}}$. 
\par As initial value for $\gamma$, we use the Hill estimator \citep{Hill} with threshold $t$ applied to $\bs{d}$.
\par This gives $\bTheta^{(0)}=(\pi^{(0)}, \bs{\beta}^{(0)}, \theta^{(0)}, \gamma^{(0)})$ where $\bs{\alpha}^{(0)}$ is transformed to $\bs{\beta}^{(0)}$ using (2).

\subsection{E-step}

In the $h$th iteration of the E-step, we take the conditional expectation of the complete log-likelihood given the incomplete data $\mathcal{X}$ and using the current estimate $\bs{\Theta}^{(h-1)}$ for $\bs{\Theta}$. As said before, we assume that there are no shared parameters in $\pi$, $\bTheta_1$ and $\bTheta_2$. We can hence split the maximisation in three parts. Therefore, we also consider three parts in the conditional expectation of the complete log-likelihood.

\subsubsection{\texorpdfstring{$\pi$}{pi}}
We easily obtain the probability (10) using (3) and (4) with $\bTheta_1=\bTheta_1^{(h-1)}$ and $\gamma=\gamma^{(h-1)}$.
This gives
\begin{align}\label{eq:prob1_MEpar}
&P\left(X_i\leq t \cond\ifive;\bTheta^{(h-1)}\right)\nonumber\\
&\qquad = \frac{\pi^{(h-1)}-\pi^{(h-1)} F_1\left(l_i; t^l, t, \bs{r}, \bTheta_1^{(h-1)}\right)}{\pi^{(h-1)}+(1-\pi^{(h-1)})\frac{1-\left(\frac{u_i}{t}\right)^{-\frac1{\gamma^{(h-1)}}}}{1-\left(\frac{T}{t}\right)^{-\frac1{\gamma^{(h-1)}}}}-\pi^{(h-1)} F_1\left(l_i; t^l, t, \bs{r}, \bTheta_1^{(h-1)}\right)},
\end{align}
where
\begin{align*}
F_1\left(l_i; t^l, t, \bs{r}, \bTheta_1^{(h-1)}\right)&=\sum_{j=1}^{M} \beta_j^{(h-1)}  F_E^t(l_i; t^l, t, r_j, \theta^{(h-1)})\\
&= \sum_{j=1}^{M} \beta_j^{(h-1)}  \frac{F_E(l_i; r_j, \theta^{(h-1)})}{F_E( t  ; r_j, \theta^{(h-1)}) -  F_E( t^l  ; r_j, \theta^{(h-1)})}.
\end{align*}

\subsubsection{ME distribution}\label{sec:app_E_ME}
Besides overcoming the estimation problem related to the censored data, as explained in Section~3, the EM algorithm also offers the right estimation framework when one of the splicing components is a mixture distribution. 
The clue is to view the data points coming from the mixture as being incomplete since the associated component-indicator vectors are not available \citep{mclachlan2000finite}.
The complete data $\mathcal{Y}$ introduced above contains the uncensored sample  $\bs{x}=(x_1,\ldots, x_{n_1}, x_{n_1+1}, \ldots, x_n)$ where we, without loss of generality, assume that the first $n_1$ data points are smaller than or equal to $t$. We further extend $\mathcal{Y}$ with component-indicator vectors for the first $n_1$ data points, denoted by $\boldsymbol{z}=(\bs{z}_{1}, \ldots , \bs{z}_{n_1})$ where
\begin{equation} z_{ij} = \begin{cases} 1 & \mbox{if $x_i$ comes from the $j$th component density }f_E^t( \cdot ;  t^l, t,  r_j, \theta) \\
	0 & \mbox{otherwise} 	\end{cases}  \label{eq:z_ij} 
\end{equation}
for $i = 1, \ldots, n_1$ and $j= 1, \ldots, M$. They are distributed according to a multinomial distribution with
\[  P ( \bs{Z}_i = \bs{z}_i; \bs{\beta}) = \beta_1^{z_{i1}} \ldots, \beta_M^{z_{iM}}  \]
for $i = 1, \ldots, n_1$,  where $z_{ij}$ is equal to 0 or 1 and $\sum_{j=1}^M z_{ij} = 1$. 
The joint density of $(X_i, \bs{Z}_{i})$ given $\{X_i \leq t\}$ equals
\begin{align*}
f_{X_i,\bs{Z}_i}(x_i, \bs{z}_i ; t^l, t, \bs{r}, \bTheta_1) &=  f_{X_i\cond\bs{Z}_i}(x_i\cond\bs{z}_i ; t^l, t, \bs{r}, \bTheta_1)\, P(\bs{Z}_i=\bs{z}_i ; \bs{\beta}) \\
&= \prod_{j=1}^{M} \left( f^t_E(x_i ; t^l, t, r_j, \theta)\right)^{z_{ij}} \prod_{j=1}^{M} \beta_j^{z_{ij}} \\
&= \prod_{j=1}^{M} \left(\beta_j f^t_E(x_i ; t^l, t, r_j, \theta)\right)^{z_{ij}} \, ,
\end{align*}
for $i = 1, \ldots, n_1$. Hence the part of the complete data log-likelihood (6) depending on $\bTheta_1$ becomes
\begin{align} 
\sum_{i=1}^n I(x_i\leq t) \ln f_1(x_i; t^l, t, \bTheta_1) = \sum_{i=1}^{n_1} \sum_{j=1}^{M} z_{ij}\ln \left(\beta_j f^t_E(x_i ; t^l, t, r_j,  \theta)\right).  \label{eq:cll_mix_erlang} 
\end{align}

\par After using arguments similar to those used to obtain (9), we get that the conditional expectation of \eqref{eq:cll_mix_erlang}, the part of the complete data log-likelihood depending on $\bTheta_1$, is given by
\begin{align}\label{eq:max_EM} 
&\sum_{i \in S_{i.}}E\left(\sum_{j=1}^{M} Z_{ij} \ln \left(\beta_j   f_E^t(X_i;  t^l, t,  r_j, \theta)\right)  \cond \ione; \bTheta_1^{(h-1)} \right) \nonumber\\ 
+&\sum_{i \in S_{iii.}}E\left( \sum_{j=1}^{M} Z_{ij} \ln \left(\beta_j   f_E^t(X_i;  t^l, t,  r_j, \theta)\right) \cond \ithree;\bTheta_1^{(h-1)}\right)  \nonumber\\ 
+&\,\sum_{i \in S_{v.}}E\left(\sum_{j=1}^{M} Z_{ij} \ln \left(\beta_j   f_E^t(X_i;  t^l, t,  r_j, \theta)\right) \cond\ifivelower;\bTheta_1^{(h-1)}\right) \nonumber\\ 
& \qquad\quad\times P\left(X_i\leq t\cond \ifive;\bTheta^{(h-1)}\right).
\end{align}

\paragraph{Uncensored data.}
Following \citet{ME} we obtain that the first part of the sum is equal to
\begin{align*}
 \sum_{i \in S_{i.}} \sum_{j=1}^{M}  \ {}^{i.}z_{ij}^{(h)} \Big[ &\ln \beta_j + (r_j-1) \ln x_i- \frac{x_i}{\theta}  - r_j \ln \theta  \nonumber\\
&   - \ln((r_j-1)!) - \ln \left( F_E( t  ; r_j, \theta) -  F_E( t^l  ; r_j, \theta) \right) \Big] \,.   
\end{align*}
Here is
\begin{align}\label{eq:zu}
{}^{i.}z_{ij}^{(h)} &:=  P\left(Z_{ij} = 1 \cond \ione;  \bTheta_1^{(h-1)} \right) = \frac{\alpha_j^{(h-1)} f_E(x_i;  r_j, \theta^{(h-1)})}{ \sum_{m=1}^{M} \alpha_m^{(h-1)} f_E(x_i; r_m, \theta^{(h-1)}) }
\end{align}
the posterior probability that an uncensored data point $x_i$  with $x_i\leq t$ (hence $i \in S_{i.}$) belongs to the $j$th component in the mixture.
In the E-step for the uncensored data, only these probabilities need to be computed for all $i\in S_{i.}$ and $j=1,\ldots, M$. They remain the same in the truncated and the untruncated case.

\paragraph{Censored data.}
We now have to distinguish between the cases \textit{iii} and \textit{v}. The derivations for case \textit{iii} follow from \citet{ME}.
\par Denote by ${}^{iii.}z_{ij}^{(h)}$ the posterior probability that the data point $x_i$ belongs to the $j$th component in the mixture for a censored data point with $u_i\leq t$.
Then,
\begin{align}
&\sum_{i \in S_{iii.}} E\left(\sum_{j=1}^{M} Z_{ij} \ln \left(\beta_j   f_E^t(X_i;  t^l, t,  r_j, \theta)\right) \cond \ithree;\bTheta_1^{(h-1)}\right) \nonumber \\
&= \sum_{i \in S_{iii.}} \sum_{j=1}^{M} {}^{iii.}z_{ij}^{(h)} E \left( \ln \left(\beta_j   f_E^t(X_i;  t^l, t,  r_j, \theta)\right)  \cond Z_{ij} = 1, \ithree;\theta^{(h-1)} \right) \nonumber \\
&= \sum_{i \in S_{iii.}}\sum_{j=1}^{M} {}^{iii.}z_{ij}^{(h)} \Bigg[ \ln \beta_j + (r_j-1) E\left(\ln X_i\cond Z_{ij} = 1, \ithree; \theta^{(h-1)} \right)   \nonumber \\
&\qquad \qquad - \frac{1}{\theta} E\left(X_i \cond Z_{ij} = 1,\ithree; \theta^{(h-1)}\right) - r_j \ln \theta- \ln((r_j-1)!) \nonumber \\
&\qquad  \qquad - \ln \left( F_E( t  ; r_j, \theta) -  F_E( t^l  ; r_j, \theta) \right) \Bigg] \label{eq:Q_c_mix_erlang}
\end{align}
where we use the law of total expectation in the first equality. 
Using Bayes' rule, we can compute the posterior probabilities  ${}^{iii.}z_{ij}^{(h)}$, for $i \in S_{iii.}$  and $j=1,\ldots, M$, as
\begin{align}
{}^{iii.}z_{ij}^{(h)} &:= P\left(Z_{ij}=1 \cond \ithree;\bTheta_1^{(h-1)}\right) \nonumber \\
&= \frac{ \beta_j^{(h-1)} \left( F_E^t(u_i;  t^l, t,  r_j, \theta^{(h-1)})  - F_E^t(l_i;  t^l, t,  r_j, \theta^{(h-1)}) \right) } { \sum_{m=1}^{M} \beta_m^{(h-1)} \left( F_E^t(u_i;  t^l, t,  r_m, \theta^{(h-1)}) - F_E^t(l_i;  t^l, t,  r_m, \theta^{(h-1)}) \right) }  \nonumber\\
& = \frac{ \alpha_j^{(h-1)} \left(  F_E( u_i  ; r_j, \theta^{(h-1)}) -  F_E( l_i  ; r_j, \theta^{(h-1)}) \right) } { \sum_{m=1}^{M} \alpha_m^{(h-1)} \left(  F_E( u_i  ; r_m, \theta^{(h-1)}) -  F_E( l_i  ; r_m, \theta^{(h-1)}) \right)  }   \, . \label{eq:c^z_ij^(h)}
\end{align}
The expression for the posterior probability in the censored case has the same form as in the uncensored case \eqref{eq:zu}, but with the densities replaced by the probabilities to lay between the lower and upper censoring points. The terms in (\ref{eq:Q_c_mix_erlang}) containing $(r_j-1)\allowbreak  E\left(\ln X_i\cond Z_{ij} = 1, \ifive; \theta^{(h-1)} \right)  $ do not play a role in the EM algorithm as they do not depend on $\bs{\beta}$ or $\theta$. We also need to compute the following conditional expected value in the E-step:
\begin{align}
&E\left( X_i \cond Z_{ij} = 1, \ithree; \theta^{(h-1)} \right)\nonumber\\
&\quad =   \int_{l_i}^{u_i} x \frac{f_E(x ; r_j , \theta^{(h-1)}) }{  F_E( u_i  ; r_j, \theta^{(h-1)}) -  F_E( l_i  ; r_j, \theta^{(h-1)}) }\, dx  \nonumber \\
& \quad =  \frac{r_j \theta^{(h-1)}}{F_E( u_i  ; r_j, \theta^{(h-1)}) -  F_E( l_i  ; r_j, \theta^{(h-1)})} \int_{l_i}^{u_i} \frac{ x^{r_j} \exp\left(-x/\theta^{(h-1)}\right) }{ \left( \theta^{(h-1)} \right)^{r_j+1} r_j!  }\,  dx    \nonumber \\
& \quad =  \frac{r_j \theta^{(h-1)} \left( F_E( u_i  ; r_j+1, \theta^{(h-1)}) -  F_E( l_i  ; r_j+1, \theta^{(h-1)}) \right) }{F_E( u_i  ; r_j, \theta^{(h-1)}) -  F_E( l_i  ; r_j, \theta^{(h-1)})}   \, ,
\label{eq:Estep_ME_c}
\end{align}
for $i \in S_{iii.}$ and $j=1,\ldots, M$, which has a closed-form expression.

\par We perform similar calculations for case \textit{v}.
 Denote by ${}^{v.}z_{ij}^{(h)}$ the posterior probability that the data point $x_i$ belongs to the $j$th component in the mixture for a censored data point with $\{X_i\leq t\}$ and $u_i> t$.
The third part of the sum in \eqref{eq:max_EM}, without the probability, is then given by
\begin{align}
&\sum_{i \in S_{v.}} E\left(\sum_{j=1}^{M} Z_{ij} \ln \left(\beta_j   f_E^t(X_i;  t^l, t,  r_j, \theta)\right) \cond \ifivelower;\bTheta_1^{(h-1)}\right) \nonumber \\
&= \sum_{i \in S_{v.}}\sum_{j=1}^{M} {}^{v.}z_{ij}^{(h)} \Bigg[ \ln \beta_j + (r_j-1) E\left(\ln X_i\cond Z_{ij} = 1,  \ifivelower; \theta^{(h-1)} \right)   \nonumber \\
&\qquad \qquad    - \frac{1}{\theta} E\left(X_i \cond Z_{ij} = 1,\ifivelower, \theta^{(h-1)}\right) - r_j \ln \theta- \ln((r_j-1)!) \nonumber \\
&\qquad \qquad    - \ln \left( F_E( t  ; r_j, \theta) -  F_E( t^l  ; r_j, \theta) \right) \Bigg]. \label{eq:Q_v_mix_erlang}
\end{align}
The posterior probabilities ${}^{v.}z_{ij}^{(h)}$, for $i \in S_{v.}$  and $j=1,\ldots, M$, are given by
\begin{align}
{}^{v.}z_{ij}^{(h)} &:= P\left(Z_{ij}=1 \cond \ifivelower;\bTheta_1^{(h-1)}\right) \nonumber \\
&=\frac{  \beta_j^{(h-1)} \left(F_E^t(t;  t^l, t,  r_j, \theta^{(h-1)}) -  F_E^t(l_i;  t^l, t,  r_j, \theta^{(h-1)}) \right) } { \sum_{m=1}^{M} \beta_m^{(h-1)} \left(F_E^t(t;  t^l, t,  r_m, \theta^{(h-1)}) - F_E^t(l_i;  t^l, t,  r_m, \theta^{(h-1)}) \right)} \nonumber\\
& = \frac{ \alpha_j^{(h-1)} \left(  F_E( t  ; r_j, \theta^{(h-1)}) -  F_E( l_i  ; r_j, \theta^{(h-1)}) \right) } { \sum_{m=1}^{M} \alpha_m^{(h-1)} \left(  F_E( t  ; r_m, \theta^{(h-1)}) -  F_E( l_i  ; r_m, \theta^{(h-1)}) \right)  }.\label{eq:{v}^z_ij^(h)}
\end{align}
This expression corresponds to \eqref{eq:c^z_ij^(h)} with $t$ instead of $u_i$ because of the conditioning on the event $\{X_i\leq t<u_i\}$.
Again, the terms in (\ref{eq:Q_v_mix_erlang}) containing $(r_j-1)\allowbreak \\ E\left(X_i \cond Z_{ij} = 1,\ifivelower; \theta^{(h-1)} \right)$ do not play a role in the EM algorithm as they do not depend on $\bs{\beta}$ or $\theta$. Also,
\begin{align}\label{eq:Estep_ME_v}
&E\left( X_i \cond Z_{ij} = 1, \ifivelower;\theta^{(h-1)} \right)\nonumber\\
&\quad =  \int_{l_i}^{t} x \frac{f_E(x ; r_j , \theta^{(h-1)}) }{  F_E( t  ; r_j, \theta^{(h-1)}) -  F_E( l_i  ; r_j, \theta^{(h-1)}) }\, dx   \nonumber \\
& \quad =  \frac{r_j \theta^{(h-1)} \left( F_E( t ; r_j+1, \theta^{(h-1)}) -  F_E( l_i  ; r_j+1, \theta^{(h-1)}) \right) }{F_E(t  ; r_j, \theta^{(h-1)}) -  F_E( l_i  ; r_j, \theta^{(h-1)})},
\end{align}
for $i \in S_{v.}$  and $j=1,\ldots, M$, which has a closed-form expression.
Likewise,  \eqref{eq:Estep_ME_v} corresponds to \eqref{eq:Estep_ME_c} with $t$ instead of $u_i$ since we condition on the event $\{X_i\leq t<u_i\}$.

\subsubsection{Pareto distribution}\label{sec:app_E_Pa}

\noindent The relevant part for the Pareto parameter $\gamma$ is
\begin{align*}
 &\sum_{i \in S_{ii.}} E\left(\ln f_2(X_i;t,T,\gamma) \cond \itwo;\gamma^{(h-1)}\right)\\
+& \sum_{i \in S_{iv.}} E\left(\ln f_2(X_i;t,T,\gamma) \cond\ifour;\gamma^{(h-1)}\right)\\
+&\, \sum_{i \in S_{v.}} E\left(\ln f_2(X_i;t,T,\gamma) \cond  \ifiveupper;\gamma^{(h-1)}\right)\\
& \qquad\  \times  P\left(X_i>t \cond  \ifive;\bTheta^{(h-1)}\right).
\end{align*}
The first conditional expectation is simply equal to
\begin{align*}
 E&\left(\ln f_2(X_i;t,T,\gamma) \cond \itwo;\gamma^{(h-1)}\right) \\
&=\ln f_2(x_i;t,T,\gamma)= -\ln(\gamma t)-\left(\frac1{\gamma}+1\right)\ln\left(\frac{x_i}{t}\right)-\ln\left(1- \left(\frac{T}{t}\right)^{-\frac1{\gamma}}\right).
\end{align*}
Starting from the definition we obtain
\begin{align*}
&E\left(\ln f_2(X_i;t,T,\gamma) \cond\ifour;\gamma^{(h-1)}\right) \\
&=-\ln(\gamma t) -\left(\frac1{\gamma}+1\right) E\left(\ln\left(\frac{X_i}t\right)\cond\ifour;\gamma^{(h-1)}\right) -\ln\left(1- \left(\frac{T}{t}\right)^{-\frac1{\gamma}}\right).
\end{align*}
Using integration by parts, we get
\begin{align}\label{eq:Estep_Par}
&E\left(\ln\left(\frac{X_i}t\right) \cond\ifour;\gamma^{(h-1)}\right)\nonumber\\
&=  \int_{l_i}^{u_i} \ln\left(\frac{x}t\right) \frac{f_2^*\left(x;t,\gamma^{(h-1)}\right)}{F_2^*\left(u_i;t,\gamma^{(h-1)}\right)-F_2^*\left(l_i;t,\gamma^{(h-1)}\right)} \, dx \nonumber\\
&=  \frac1{\left(\frac{l_i}{t}\right)^{-\frac1{\gamma^{(h-1)}}}-\left(\frac{u_i}{t}\right)^{-\frac1{\gamma^{(h-1)}}}}\int_{l_i}^{u_i} \frac{\ln\left(\frac{x}t\right)}{\gamma^{(h-1)}t} \left(\frac{x}{t}\right)^{-\frac1{\gamma^{(h-1)}}-1} \, dx\nonumber\\
&=  \frac1{\left(\frac{l_i}{t}\right)^{-\frac1{\gamma^{(h-1)}}}-\left(\frac{u_i}{t}\right)^{-\frac1{\gamma^{(h-1)}}}}\int_{\frac{l_i}{t}}^{\frac{u_i}{t}} \frac{\ln v}{\gamma^{(h-1)}} v^{-\frac1{\gamma^{(h-1)}}-1} \, dv\nonumber\\
&=  \frac1{\left(\frac{l_i}{t}\right)^{-\frac1{\gamma^{(h-1)}}}-\left(\frac{u_i}{t}\right)^{-\frac1{\gamma^{(h-1)}}}}\left(-\left[\ln v\  v^{-\frac1{\gamma^{(h-1)}}} \right]_{\frac{l_i}{t}}^{\frac{u_i}{t}}+  \int_{\frac{l_i}{t}}^{\frac{u_i}{t}}v^{-\frac1{\gamma^{(h-1)}}-1}  \,dv\right)\nonumber\\
&=  \frac1{\left(\frac{l_i}{t}\right)^{-\frac1{\gamma^{(h-1)}}}-\left(\frac{u_i}{t}\right)^{-\frac1{\gamma^{(h-1)}}}}\left(-\left[\ln v \ v^{-\frac1{\gamma^{(h-1)}}} \right]_{\frac{l_i}{t}}^{\frac{u_i}{t}} + \left[-\gamma^{(h-1)}v^{-\frac1{\gamma^{(h-1)}}} \right]_{\frac{l_i}{t}}^{\frac{u_i}{t}}\right)\nonumber\\
&=\frac{\left(\ln\left(\frac{l_i}t\right)+\gamma^{(h-1)}\right)\left(\frac{l_i}{t}\right)^{-\frac1{\gamma^{(h-1)}}} - \left(\ln\left(\frac{u_i}t\right)+\gamma^{(h-1)}\right)\left(\frac{u_i}{t}\right)^{-\frac1{\gamma^{(h-1)}}}}{\left(\frac{l_i}{t}\right)^{-\frac1{\gamma^{(h-1)}}}-\left(\frac{u_i}{t}\right)^{-\frac1{\gamma^{(h-1)}}}}.
\end{align}
We compute the third conditional expectation similarly which gives
\begin{align*}\label{eq:Estep_Par2}
&E\Big(\left.\ln f_2(X_i;t,T,\gamma) \cond \ifiveupper;\gamma^{(h-1)}\right)\\
&=-\ln(\gamma t) -\left(\frac1{\gamma}+1\right) E\left(\ln\left(\frac{X_i}t\right) \cond\ifiveupper;\gamma^{(h-1)}\right) -\ln\left(1- \left(\frac{T}{t}\right)^{-\frac1{\gamma}}\right),
\end{align*}
with
\begin{align}
E\left(\ln\left(\frac{X_i}t\right) \cond\ifiveupper;\gamma^{(h-1)}\right)=  \frac{\gamma^{(h-1)} - \left(\ln\left(\frac{u_i}t\right)+\gamma^{(h-1)}\right)\left(\frac{u_i}{t}\right)^{-\frac1{\gamma^{(h-1)}}}}{1-\left(\frac{u_i}{t}\right)^{-\frac1{\gamma^{(h-1)}}}}.
\end{align}
Note that the last expression corresponds to \eqref{eq:Estep_Par} with $l_i=t$ since we condition on the event $\{l_i\leq t<X_i\}$. 

\subsection{M-step}

In the M-step, the expected value of the complete data log-likelihood obtained in the E-step is maximised w.r.t.~the parameter vector $\bs{\Theta}$.
As said before, we assume that there are no shared parameters in $\pi$, $\bTheta_1$ and $\bTheta_2$. This assures that we can split the maximisation in three parts.

\subsubsection{\texorpdfstring{$\pi$}{pi}}
Maximisation with respect to $\pi$ gives
\[\pi^{(h)}=\frac{n_1^{(h)}}n=\frac{\# S_{i.}+ \#S_{iii.}+\sum_{i \in S_{v.}} P\left(X_i\leq t \cond  \ifive;\bTheta^{(h-1)}\right)}n,\]
see (11).
An expression for the probability can be found in \eqref{eq:prob1_MEpar}.

\subsubsection{ME distribution}
The expected value of the complete data log-likelihood obtained in the E-step is now maximised with respect to the parameter vector $\bs{\Theta}_1=(\bs{\beta},\theta)$ over all $(\bs{\beta},\theta)$ with $\beta_j>0$, $\sum_{j=1}^{M} \beta_j=1$ and $\theta>0$. 

\par To maximise over the mixing weights $\bs{\beta}$, 
\begin{align*}
&\sum_{j=1}^{M}\left(\sum_{i \in S_{i.}}  {}^{i.}z_{ij}^{(h)} + \sum_{i \in S_{iii.}} {}^{iii.}z_{ij}^{(h)} + \sum_{i \in S_{v.}} P_{1,i}^{(h)} \ {}^{v.}z_{ij}^{(h)}\right)\ln \beta_j
\end{align*}
needs to be maximised where $P_{1,i}^{(h)} =P\left(X_i\leq t\cond \ifive;\bTheta^{(h-1)}\right)$, see \eqref{eq:prob1_MEpar}.
Setting $\beta_M = 1- \sum_{j=1}^{M-1}\beta_j$ makes sure that the restriction $\sum_{j=1}^{M} \beta_j = 1$ holds.
Equating the partial derivatives at $\bs{\beta}^{(h)}$ to zero gives
\begin{align*}
\beta_j^{(h)} =  \frac{\sum_{i \in S_{i.}}  {}^{i.}z_{ij}^{(h)} + \sum_{i \in S_{iii.}} {}^{iii.}z_{ij}^{(h)} + \sum_{i \in S_{v.}}  P_{1,i}^{(h)} \ {}^{v.}z_{ij}^{(h)}}{\sum_{i \in S_{i.}}  {}^{i.}z_{iM}^{(h)} + \sum_{i \in S_{iii.}} {}^{iii.}z_{iM}^{(h)} + \sum_{i \in S_{v.}}  P_{1,i}^{(h)} \ {}^{v.}z_{iM}^{(h)}} \beta_M^{(h)}
\end{align*}
for $j=1,\ldots, M-1$. Applying the sum constraint gives
\[\beta_M^{(h)} = \frac{\sum_{i \in S_{i.}}  {}^{i.}z_{iM}^{(h)} + \sum_{i \in S_{iii.}} {}^{iii.}z_{iM}^{(h)} + \sum_{i \in S_{v.}}  P_{1,i}^{(h)} \ {}^{v.}z_{iM}^{(h)}}{n_1^{(h)}}.
\]
The same form also follows for $j=1,\ldots, M-1$:
\begin{equation}
\beta_j^{(h)} = \frac{\sum_{i \in S_{i.}}  {}^{i.}z_{ij}^{(h)} + \sum_{i \in S_{iii.}} {}^{iii.}z_{ij}^{(h)} + \sum_{i \in S_{v.}}  P_{1,i}^{(h)} \ {}^{v.}z_{ij}^{(h)}}{n_1^{(h)}}.
\label{eq:M_step_beta}
\end{equation}
The estimate for the prior probability $\beta_j$ in the truncated mixture is thus the average of the posterior probabilities of belonging to the $j$th component in the mixture.
\par In order to maximise with respect to $\theta$, we set the first order partial derivative at $\theta^{(h)}$ equal to zero
\begin{align*}
0&=\sum_{j=1}^{M}\Bigg(\sum_{i \in S_{i.}} {}^{i.}z_{ij}^{(h)} \left(  -\frac{r_j}{\theta} - \frac{ \frac{\partial }{\partial \theta}\left[ F_E( t  ; r_j, \theta) -  F_E( t^l  ; r_j, \theta) \right]}{ F_E( t  ; r_j, \theta) -  F_E( t^l  ; r_j, \theta) } + \frac{x_i}{\theta^2}\right)   \\
& \qquad\qquad + \sum_{i \in S_{iii.}} {}^{iii.}z_{ij}^{(h)} \left(   -\frac{r_j}{\theta}- \frac{ \frac{\partial }{\partial \theta}\left[ F_E( t  ; r_j, \theta) -  F_E( t^l  ; r_j, \theta) \right]}{ F_E( t  ; r_j, \theta) -  F_E( t^l  ; r_j, \theta) }\right. \\
& \qquad \qquad\qquad\quad+\left. \frac{E\left(X_i \cond Z_{ij} = 1, \ithree ; \theta^{(h-1)} \right)}{\theta^2}\right) \\
&  \qquad\qquad+\, \sum_{i \in S_{v.}} P_{1,i}^{(h)} \ {}^{v.}z_{ij}^{(h)}  \Bigg( -\frac{r_j}{\theta}- \frac{ \frac{\partial }{\partial \theta}\left[ F_E( t  ; r_j, \theta) -  F_E( t^l  ; r_j, \theta) \right]}{ F_E( t  ; r_j, \theta) -  F_E( t^l  ; r_j, \theta) }  \\
& \qquad \qquad\qquad\quad+ \left.\frac{E\left(X_i \cond Z_{ij} = 1, \ifivelower; \theta^{(h-1)} \right)}{\theta^2} \Bigg)  \Bigg)\right|_{\theta = \theta^{(h)}}\\
&= -\frac{1}{\theta^{(h)}}\sum_{j=1}^{M} \Bigg(\sum_{i \in S_{i.}} {}^{i.}z_{ij}^{(h)} + \sum_{i \in S_{iii.}} {}^{iii.}z_{ij}^{(h)} + \sum_{i \in S_{v.}} P_{1,i}^{(h)} \ {}^{v.}z_{ij}^{(h)} \Bigg)r_j\\
&\quad -\sum_{j=1}^{M} \Bigg(\sum_{i \in S_{i.}} {}^{i.}z_{ij}^{(h)} +\sum_{i \in S_{iii.}} {}^{iii.}z_{ij}^{(h)} + \sum_{i \in S_{v.}} P_{1,i}^{(h)} \ {}^{v.}z_{ij}^{(h)} \Bigg) \left.\frac{ \frac{\partial }{\partial \theta}\left[ F_E( t  ; r_j, \theta) -  F_E( t^l  ; r_j, \theta) \right]}{ F_E( t  ; r_j, \theta) -  F_E( t^l  ; r_j, \theta) } \right|_{\theta = \theta^{(h)}}\\
&\quad+\frac1{\theta^{(h)2}} \sum_{j=1}^{M} \Bigg(\sum_{i \in S_{i.}} {}^{i.}z_{ij}^{(h)}\ x_i\\
&\qquad\qquad\qquad+\sum_{i \in S_{iii.}}{}^{iii.}z_{ij}^{(h)}  E\left(X_i \cond Z_{ij} = 1,  \ithree ; \theta^{(h-1)} \right) \\
&\qquad\qquad\qquad+\, \sum_{i \in S_{v.}} P_{1,i}^{(h)} \ {}^{v.}z_{ij}^{(h)}E\left(X_i \cond Z_{ij} = 1, \ifivelower; \theta^{(h-1)} \right)\Bigg). 
\end{align*}
Using the lower incomplete gamma function $\displaystyle \gamma(s,x) = \int_{0}^{x} z^{s-1}\exp(-z)\, dz$ we write the cumulative distribution function of an Erlang distribution as
\[ F_E(x; r_j, \theta) =   \int_{0}^{x}  \frac{z^{r-1} \exp(-z/\theta)}{\theta^{r_j}(r_j-1)!}\, dz  = \frac{1}{(r-1)!} \int_{0}^{x/\theta}  u^{r_j-1} \exp(-u)\, du = \frac{\gamma(r_j,x/\theta)}{(r_j-1)!} \, . \]
Applying the Leibniz rule gives that the partial derivative of $F_E$ with respect to $\theta$ is equal to
\[\frac{\partial F_E(x; r_j, \theta)}{\partial \theta}= \frac{\frac{\partial \gamma(r_j,x/\theta)}{\partial \theta}}{(r_j-1)!}=\frac{\left(\frac x{\theta}\right)^{r_j-1}\exp\left(-\frac x{\theta}\right)\left(-\frac{x}{\theta^2}\right)}{(r_j-1)!}.\]
Using \eqref{eq:M_step_beta} and this derivative gives
\begin{align*}
0 &=  -\frac{n_1^{(h)}}{\theta^{(h)}} \sum_{j=1}^{M} \beta_j^{(h)}r_j -n_1^{(h)} \sum_{j=1}^{M} \beta_j^{(h)} \left.\frac{\left(\frac{t^l}{\theta}\right)^{r_j-1}\exp\left(-\frac{t^l}{\theta}\right)\frac{t^l}{\theta^2}-\left(\frac t{\theta}\right)^{r_j-1}\exp\left(-\frac{t}{\theta}\right)\frac{t}{\theta^2}}{(r_j-1)! \left( F_E( t  ; r_j, \theta) -  F_E( t^l  ; r_j, \theta) \right)} \right|_{\theta = \theta^{(h)}}\\
&\quad+\frac1{\theta^{(h)2}} \sum_{j=1}^{M} \Bigg(\sum_{i \in S_{i.}} {}^{i.}z_{ij}^{(h)}\ x_i\\
&\qquad\qquad\qquad+\sum_{i \in S_{iii.}} {}^{iii.}z_{ij}^{(h)}  E\left(X_i \cond Z_{ij} = 1,  \ithree ; \theta^{(h-1)}\right) \\
&\qquad\qquad\qquad+\, \sum_{i \in S_{v.}} P_{1,i}^{(h)} \ {}^{v.}z_{ij}^{(h)}E\left(X_i \cond Z_{ij} = 1, \ifivelower; \theta^{(h-1)} \right)\Bigg). 
\end{align*}
This leads to the following M-step equation for $\theta$:
\begin{align}\label{eq:M_step_theta}
\theta^{(h)} &= \frac{1}{n_1^{(h)}\sum_{j=1}^{M} \beta_j^{(h)}r_j} \Bigg(\sum_{i \in S_{i.}} x_i\nonumber\\
& \qquad\qquad\qquad \ + \sum_{i \in S_{iii.}}\sum_{j=1}^{M} {}^{iii.}z_{ij}^{(h)} E\left(X_i \cond Z_{ij} = 1,  \ithree ; \theta^{(h-1)} \right)\nonumber\\
& \qquad\qquad\qquad \ +\,  \sum_{i \in S_{v.}}\sum_{j=1}^{M}  P_{1,i}^{(h)} \ {}^{v.}z_{ij}^{(h)} E\left(X_i \cond Z_{ij} = 1, \ifivelower; \theta^{(h-1)} \right) \nonumber\Bigg)\\
& \ -\frac{ \sum_{j=1}^{M} \beta_j^{(h)} \frac{\left(t^l\right)^{r_j}\exp\left(-\frac{t^l}{\theta^{(h)}}\right)-t^{r_j}\exp\left(-\frac{t}{\theta^{(h)}}\right)}{\left(\theta^{(h)}\right)^{r_j-1}(r_j-1)!\left( F_E( t  ; r_j, \theta^{(h)}) -  F_E( t^l  ; r_j, \theta^{(h)}) \right)}}{\sum_{j=1}^{M} \beta_j^{(h)}r_j},
\end{align}
where expressions for the conditional expectations can be found in \eqref{eq:Estep_ME_c} and \eqref{eq:Estep_ME_v}.
This equation can be seen as the sample mean of all data points that are smaller than or equal to $t$, divided by the average shape parameter. For the censored data points, the sample mean is replaced by the expected value given the lower and upper bounds (second and third terms). Moreover, there is a correction for truncation (fourth term) which depends on $\theta^{(h)}$ itself in a complicated way. Therefore, it is not possible to find an analytical solution, and we solve \eqref{eq:M_step_theta} numerically using a Newton-type algorithm with the previous estimate $\theta^{(h-1)}$ as the starting value.
\par The expression for $\beta_j$ and the M-step equation for $\theta$ are similar to the ones in \citet{ME} where an extra term (third term) is included for data points of type \textit{v}.
\par Then, we transform the estimates for $\beta_j$ to estimates for $\alpha_j$ using \\$\widetilde{\alpha}_j/\sum_{j=1}^M \widetilde{\alpha}_j$ with
\[\widetilde{\alpha}_j = \frac{\hat{\beta}_j}{F_E(t;r_j, \hat{\theta})-F_E(t^l;r_j, \hat{\theta})},\]
and $\hat{\beta}_j$ and $\hat{\theta}$ the estimates obtained from the final EM-step.

\subsubsection{Pareto distribution}
To maximise the expected log-likelihood with respect to $\bTheta_2=\gamma$, we have to maximise
\begin{align}
\label{eq:Mstep_Pareto}
&\sum_{i \in S_{ii.}} E\left(\ln f_2(X_i;t,T,\gamma) \cond\itwo;\gamma^{(h-1)}\right) \nonumber\\
&+\sum_{i \in S_{iv.}} E\left(\ln f_2(X_i;t,T,\gamma) \cond \ifour;\gamma^{(h-1)}\right)\nonumber\\
&+\,\sum_{i \in S_{v.}} \ P_{2,i}^{(h)} E\left(\ln f_2(X_i;t,T,\gamma)\cond \ifiveupper;\gamma^{(h-1)}\right)\nonumber\\
=&\sum_{i \in S_{ii.}} \Big[-\ln(\gamma t)-\ln\left(1- \left(\frac{T}{t}\right)^{-\frac1{\gamma}}\right)-\left(\frac1{\gamma}+1\right)\ln\left(\frac{x_i}{t}\right)\Big]\nonumber\\
&+\sum_{i \in S_{iv.}}\Bigg[-\ln(\gamma t) -\ln\left(1- \left(\frac{T}{t}\right)^{-\frac1{\gamma}}\right) \nonumber\\
& \qquad \qquad- \left(\frac1{\gamma}+1\right)E\left(\ln\left(\frac{X_i}t\right) \cond\ifour;\gamma^{(h-1)}\right)\Bigg]\nonumber\\
&+\,\sum_{i \in S_{v.}}  P_{2,i}^{(h)} \Bigg[-\ln(\gamma t) -\ln\left(1- \left(\frac{T}{t}\right)^{-\frac1{\gamma}}\right) \nonumber\\
& \qquad \qquad\qquad- \left(\frac1{\gamma}+1\right)
E\left(\ln\left(\frac{X_i}t\right) \cond\ifiveupper;\gamma^{(h-1)}\right)\Bigg],
\end{align}
where $ P_{2,i}^{(h)} =P\left(X_i> t\cond \ifive;\bTheta^{(h-1)}\right)$.
Taking the derivative of \eqref{eq:Mstep_Pareto}  with respect to $\gamma$ gives
\begin{align*}
&-\frac1{\gamma} \left(\# S_{ii.}+\#S_{iv.}+\sum_{i \in S_{v.}}P_{2,i}^{(h)}\right)\\
&+\frac{1}{\gamma^2}\left(\# S_{ii.}+\#S_{iv.}+\sum_{i \in S_{v.}} P_{2,i}^{(h)}\right) \frac{\ln\left(\frac{T}{t}\right)\left(\frac{T}{t}\right)^{-\frac1{\gamma}}}{1-\left(\frac{T}{t}\right)^{-\frac1{\gamma}}}\\
& +\frac1{\gamma^2}\Bigg(\sum_{i \in S_{ii.}} \ln\left(\frac{x_i}{t}\right) + \sum_{i \in S_{iv.}}E\left(\ln\left(\frac{X_i}t\right) \cond\ifour;\gamma^{(h-1)}\right)\\
& \qquad\quad +\sum_{i \in S_{v.}}P_{2,i}^{(h)}\ E\left(\ln\left(\frac{X_i}t\right) \cond\ifiveupper;\gamma^{(h-1)}\right)\Bigg).
\end{align*}
\newpage \noindent Setting this derivative at $\gamma^{(h)}$ equal to 0 and then solving for $\gamma^{(h)}$ results in
\begin{align*}
\gamma^{(h)} = &\frac{1}{n_2^{(h)}}\Bigg( \sum_{i \in S_{ii.}}\ln\left(\frac{x_i}{t}\right)\\
&\quad +\sum_{i \in S_{iv.}}E\left(\ln\left(\frac{X_i}t\right) \cond\ifour;\gamma^{(h-1)}\right)
\\
& \quad +\,\sum_{i \in S_{v.}}P_{2,i}^{(h)}\ E\left(\ln\left(\frac{X_i}t\right) \cond\ifiveupper;\gamma^{(h-1)}\right)\Bigg)\\
&+\frac{\ln\left(\frac{T}{t}\right)}{\left(\frac{T}{t}\right)^{\frac1{\gamma^{(h)}}}-1}.
\end{align*}
where expressions for the conditional expectations can be found in \eqref{eq:Estep_Par} and \eqref{eq:Estep_Par2}.
Note that $P_{2,i}^{(h)} =1-P_{1,i}^{(h)}$ and we compute the latter using \eqref{eq:prob1_MEpar}.
The first term in the equation can be seen as the sample mean of all $\ln(X_i/t)$ with $\{X_i>t\}$ which are observed for data points belonging to case \textit{ii} and have to be replaced by their expected values for cases \textit{iv} and \textit{v}. 
Similarly to \eqref{eq:M_step_theta}, the last term, which depends on $\gamma^{(h)}$, is a correction term for the upper truncation at point $T$ and requires us to solve this equation numerically. 
In case there is no upper truncation, i.e.~$T=+\infty$, the last term is equal to zero and we obtain an analytical solution for $\gamma^{(h)}$.

\subsection{Choice of shape parameters and number of mixtures for ME distribution}\label{sec:ME_adj_red}

To choose the shape parameters $\bs{r}$ and the number of mixtures $M$, we follow the approach from \citet{MME} which is a slightly modified version of the approach from \citet{ME}.
Starting from a certain value for $M$ and shapes as given in \eqref{eq:rinit}, they try to reduce $M$ using a backward stepwise search where the mixture component with the smallest shape is deleted if this decreases an IC (AIC or BIC). 
Moreover, after the first reduction of $M$, $M$ is further reduced using the IC and the shapes $\bs{r}$ are adjusted based on maximising the likelihood.
For each value for $M$ and $\bs{r}$, the EM algorithm described above is executed.
We refer to Section~4 in \citet{MME} for more details. Important is that we now consider the \mbox{(log-)}likelihood including the Pareto part, as given in (5).

\section{Fitting the ME-Pareto model to uncensored data using the EM algorithm}\label{sec:nocens}

In case there are no censored data points, calculations are much simpler. Now, we only have data points from cases \textit{i} and \textit{ii}.

\subsection{Starting values}
We use the starting values for $\theta$ and $\bs{\beta}$ from \citet{MME}, see Section~\ref{sec:app_initial}.
We do not need starting values for $\pi$ and $\gamma$ as the obtained estimates are the same in each step.

\subsection{\texorpdfstring{$\pi$}{pi}}
Maximisation w.r.t.~$\pi$ gives
\begin{equation}\label{eq:pi_Mstep_nocens}
\pi^{(h)}=\frac{\#S_{i.}}n=\frac{n_1}n.
\end{equation}
The estimate for $\pi$ is thus equal to the proportion of points that is smaller than or equal to the splicing point $t$.
It is clear that $\pi^{(h)}$ is constant since it is independent of $h$.
\par Note that taking the splicing point equal to the $(k+1)$th largest data point, i.e.~$t=x_{n-k,n}$, gives $\displaystyle \pi^{(h)}=1-\frac{k}{n}$ when there are no ties.

\subsection{ME distribution}

Similar to before, the following expression is obtained, for $j=1,\ldots, M$,
\begin{equation*}
\beta_j^{(h)} = \frac{\sum_{i \in S_{i.}} {}^{i.}z_{ij}^{(h)}}{n_1}.
\label{eq:M_step_beta_nocens}
\end{equation*}
\par The M-step equation for $\theta$ now simplifies to
\begin{equation}\label{eq:M_step_theta_nocens}
\theta^{(h)} = \frac{\sum_{i \in S_{i.}} x_i}{n_1\sum_{j=1}^{M} \beta_j^{(h)}r_j} -\frac{ \sum_{j=1}^{M} \beta_j^{(h)} \frac{\left(t^l\right)^{r_j}\exp\left(-\frac{t^l}{\theta^{(h)}}\right)-t^{r_j}\exp\left(-\frac{t}{\theta^{(h)}}\right)}{\left(\theta^{(h)}\right)^{r_j-1}(r_j-1)!\left( F_E( t  ; r_j, \theta^{(h)}) -  F_E( t^l  ; r_j, \theta^{(h)}) \right)}}{\sum_{j=1}^{M} \beta_j^{(h)}r_j}.
\end{equation}
As before, a Newton-type algorithm is employed to solve (\ref{eq:M_step_theta_nocens}) numerically using the previous estimate $\theta^{(h-1)}$ as starting value.

\subsection{Pareto distribution}
For the Pareto parameter $\gamma$ we get following implicit equation
\begin{align}\label{eq:gamma_nocens_trunc}
\gamma^{(h)} &= \frac{\sum_{i \in S_{ii.}}  \ln\left(\frac{x_i}{t}\right)}{n_2}+\frac{\ln\left(\frac{T}{t}\right)}{\left(\frac{T}{t}\right)^{\frac1{\gamma^{(h)}}}-1}.
\end{align}
It is immediately clear that this estimate does not depend on $h$, so we denote $\hat{\gamma}=\gamma^{(h)}$. The first term of \eqref{eq:gamma_nocens_trunc} is the Hill estimator \citep{Hill} with threshold $t$, and the second term is a correction for the upper truncation at $T$ which vanishes for $T = + \infty$. Setting $t$ equal to the $(k+1)$th largest data point $x_{n-k,n}$ gives
\begin{align}\label{eq:gamma_nocens_trunc_nk}
\hat{\gamma} = \frac{1}{k}\sum_{i=1}^k \ln\left(\frac{x_{n-i+1,n}}{x_{n-k,n}}\right)+\frac{\ln\left(\frac{T}{x_{n-k,n}}\right)}{\left(\frac{T}{x_{n-k,n}}\right)^{\frac1{\hat{\gamma}}}-1}=H_{k,n}+\frac{\ln\left(\frac{T}{x_{n-k,n}}\right)}{\left(\frac{T}{x_{n-k,n}}\right)^{\frac1{\hat{\gamma}}}-1}.
\end{align}

\subsection{Upper endpoint estimation} \label{sec:upper}

\citet{trHill} show that the conditional MLE for the endpoint $T$, if it is unknown, is given by the maximum $x_{n,n}$. The corresponding conditional MLE for $\gamma$ is given by
\begin{align*}
\hat{\gamma} =H_{k,n}+\frac{\ln\left(\frac{x_{n,n}}{x_{n-k,n}}\right)}{\left(\frac{x_{n,n}}{x_{n-k,n}}\right)^{\frac1{\hat{\gamma}}}-1}.
\end{align*}%
This estimator corresponds to \eqref{eq:gamma_nocens_trunc_nk} where $T$ is estimated by its conditional MLE $x_{n,n}$. 
\citet{Truncation} use the same estimator $\hat{\gamma}$, but re-estimate the endpoint $T$ as
\[\hat{T} = \max\left\{x_{n-k,n}\left(1+\frac{k}{n \hat{D}_T}\right)^{\hat{\gamma}}, \; x_{n,n}\right\},\]
where 
\[\hat{D}_T = \max\left\{\frac kn\frac{\left(\frac{x_{n-k,n}}{x_{n,n}}\right)^{\frac1{\hat{\gamma}}}-\frac{1}{k+1}}{1-\left(\frac{x_{n-k,n}}{x_{n,n}}\right)^{\frac1{\hat{\gamma}}}}, \; 0\right\}\]
is an estimator for the truncation odds $D_T=\frac{1-F_2^*(T)}{F_2^*(T)}$.

\section{Additional QQ-plots}

\begin{figure}[ht]
			\centering
			\includegraphics[height=0.49\textwidth, angle=270]{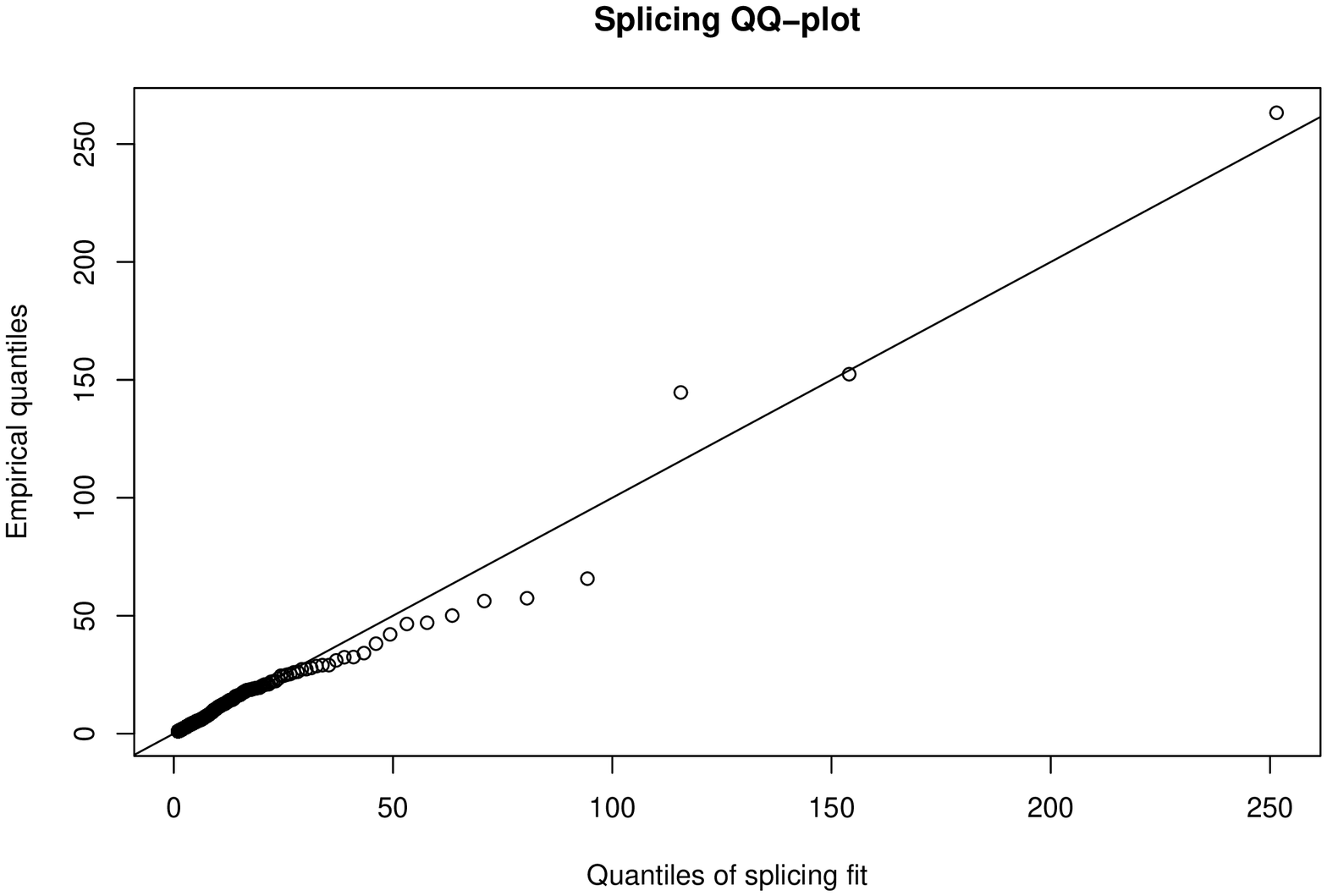}%
		~
			 \includegraphics[height=0.49\textwidth, angle=270]{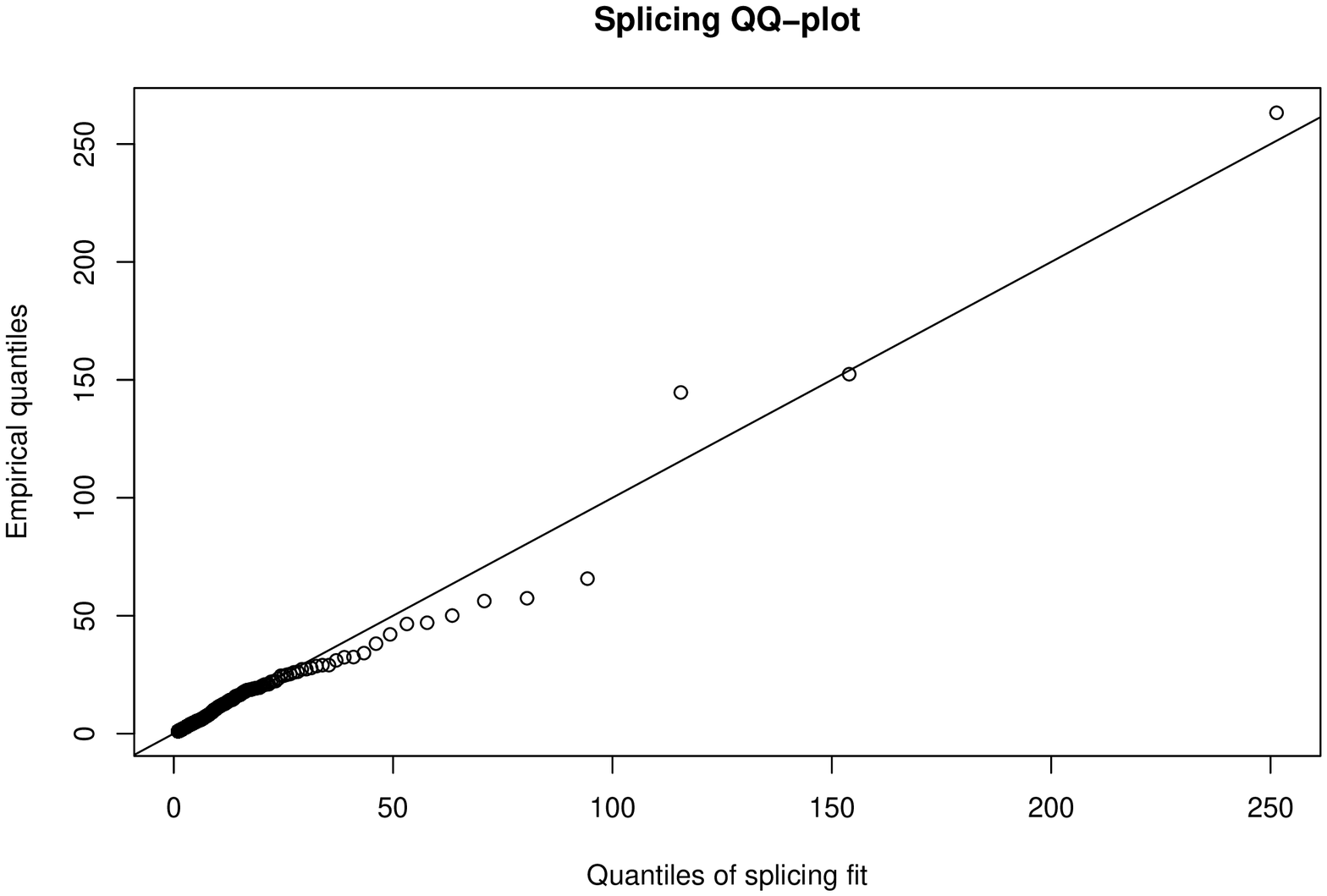}%
		\caption{Danish fire insurance: QQ-plots of the fitted LN-Pa (left) and W-Pa (right) splicing models.}%
		\label{fig:Danish_QQ_adhoc}
	\end{figure}

\begin{figure}[ht]
			\centering
			\vspace{-1cm}
			\includegraphics[height=0.49\textwidth, angle=270]{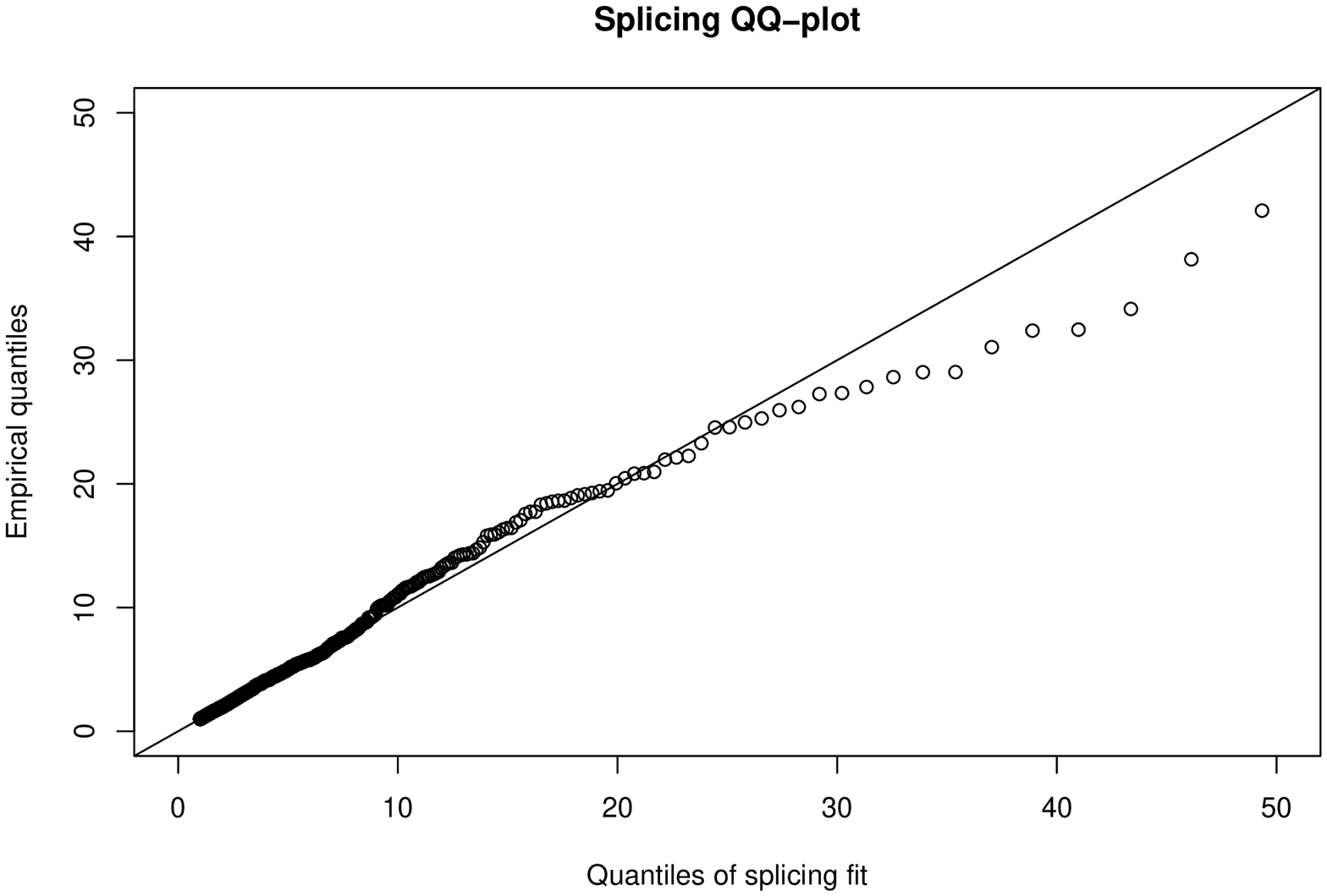}%
		~
			 \includegraphics[height=0.49\textwidth, angle=270]{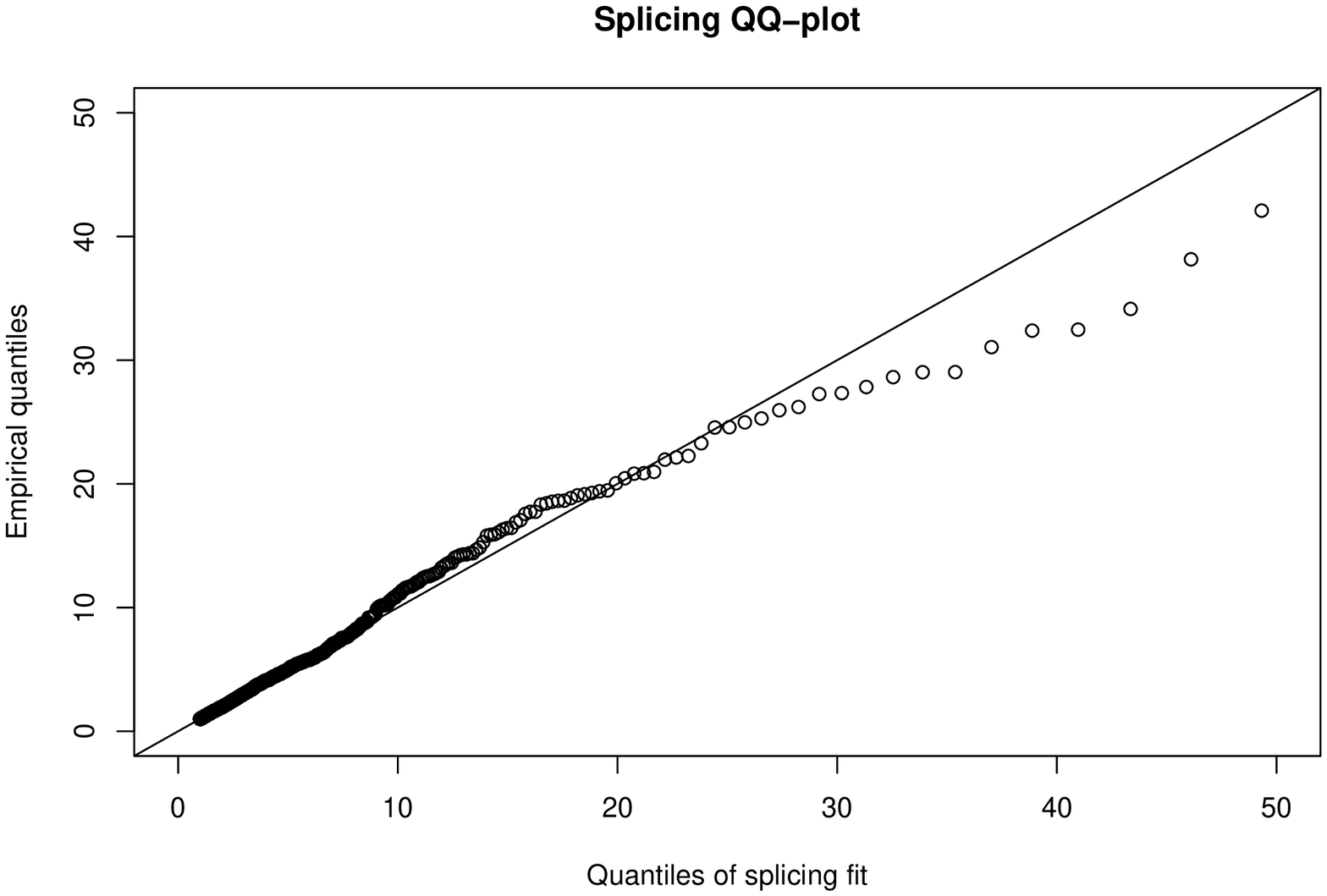}%
			\\
						 \includegraphics[height=0.49\textwidth, angle=270]{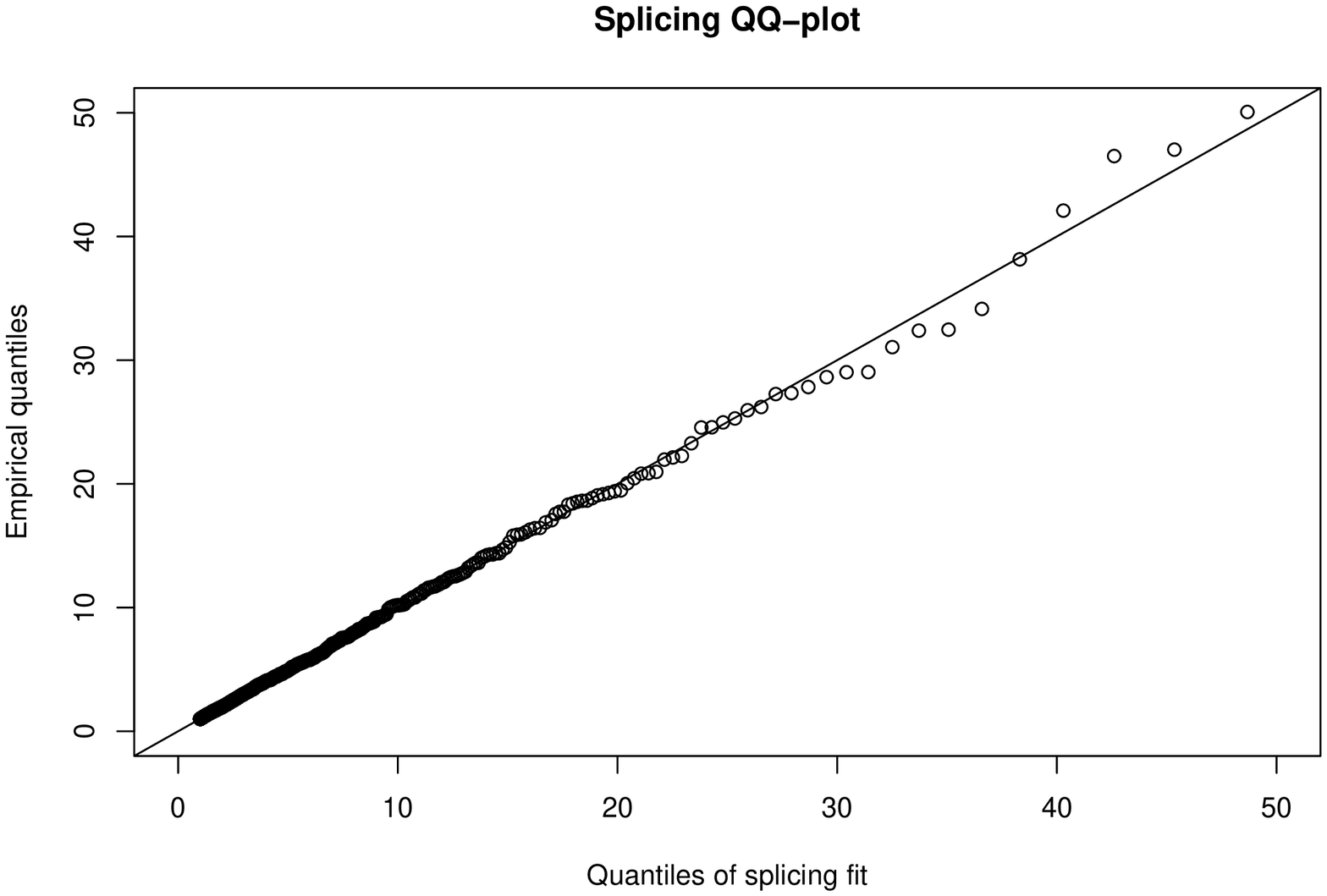}%
		\caption{Danish fire insurance: zoomed QQ-plots of the fitted LN-Pa (top left), W-Pa (top right) and ME-Pa (bottom) splicing models.}%
				\label{fig:Danish_QQ_zoom}
	\end{figure}

\bibliography{bib_splicing}

\addcontentsline{toc}{section}{References}